\documentclass[12pt]{article}
\usepackage{epsfig,a4}

\def\doublespace{\def\baselinestretch{1.6}\large\normalsize}
\def\normalspace{\def\baselinestretch{1.0}\normalsize}
\def\Address#1#2{$^{\rm#1}${\it\footnotesize#2}\\}
\def\Ref#1{(\ref{#1})}
\def\Caption#1{
  \normalspace
  \begin{quotation}\caption{\sl #1}\end{quotation}
  \doublespace
}

\def\BA{\begin{eqnarray}}
\def\BE{\begin{equation}}
\def\BF{\begin{figure}[htb]}
\def\BT{\begin{table}[htb]}
\def\EA{\end{eqnarray}}
\def\EE{\end{equation}}
\def\EF{\end{figure}}
\def\ET{\end{table}}
\def\vr{\vec r\,} \def\br{{\vec r\,}} \def\rT{r_T} \def\brT{{\vec r_T}}
\def\vp{\vec p\,}  \def\pT{p_T} \def\bpT{{\vec p_T}}
  \def\qT{q_T} 
\def\la{\langle}
\def\ra{\rangle}
\def\mb{\,\mbox{mb}}
\def\fm{\,\mbox{fm}}
\def\GeV{\,\mbox{GeV}}
\def\eps{\varepsilon}

\def\Jpsi{J\!/\!\psi}

\begin{document}

\title{ {\bf Photoproduction of Charmonia and\\ Total Charmonium-Proton
Cross Sections}}
\author{
  J.~H\"ufner$^{a,b}$,
  Yu.P.~Ivanov$^{a,b,c}$,
  B.Z.~Kopeliovich$^{b,c}$
  and A.V.Tarasov$^{a,b,c}$
  \\
  \\
  \Address{a}{
    Institut f\"ur Theoretische Physik der Universit\"at, 
    Philosophenweg 19, 69120 Heidelberg, Germany
  }
  \Address{b}{
    Max-Planck Institut f\"ur Kernphysik,
    Postfach 103980, 69029 Heidelberg, Germany
  }
  \Address{c}{
    Joint Institute for Nuclear Research,
    Dubna, 141980 Moscow Region, Russia
  }
}
\date{\today}
\maketitle
\doublespace

\begin{abstract}

Elastic virtual photoproduction cross sections $\gamma^*p\to\Jpsi(\psi')\,p$
and total charmonium-nucleon cross sections for $\Jpsi,\ \psi'$ and $\chi$
states are calculated in a parameter free way with the light-cone dipole
formalism and the same input: factorization in impact parameters, light-cone
wave functions for the $\gamma^*$ and the charmonia, and the universal
phenomenological dipole cross section which is fitted to other data. The
charmonium wave functions are calculated with four known realistic potentials,
and two models for the dipole cross section are tested. Very good agreement
with data for the cross section of charmonium photoproduction is found in
a wide range of $s$ and $Q^2$. The inclusion of the Melosh spin rotation
increases the $\psi'$ photoproduction rate by a factor $2-3$ and
removes previously observed discrepancies in the $\psi'$ to $\Jpsi$ ratio
in photoproduction. We also calculate the charmonium-proton cross sections
whose absolute values and energy dependences are found to correlate strongly
with the sizes of the states.

\end{abstract}

\newpage

\section{Introduction}\label{intro}

The dynamics of production and interaction of charmonia has drawn attention
since their discovery back in 1973. As these heavy mesons have a small size
it has been expected that hadronic cross sections may be calculated relying
on perturbative QCD. The study of charmonium production became even more
intense after charmonium suppression had been suggested as a probe for the
creation and interaction of quark-gluon plasma in relativistic heavy ion
collisions \cite{Satz}.

Since we will never have direct experimental information on charmonium-nucleon
total cross sections one has to extract it from other data for example from
elastic photoproduction of charmonia $\gamma p \to \Jpsi(\psi')\ p\,$. The
widespread believe that one can rely on the vector dominance model (VDM)
is based on previous experience the with photoproduction of $\rho$ mesons.
However, even a dispersion approach shows that this is quite a risky way,
because the $\Jpsi$ pole in the complex $Q^2$ plane is nearly 20 times
farther away from the physical region than the $\rho$ pole. The multichannel
analysis performed in \cite{HK} demonstrates that the corrections are huge,
$\sigma^{\Jpsi\,p}_{tot}$ turns out to be more that three times larger than
the VDM prediction. Unfortunately, more exact predictions of the multichannel
approach, especially for $\psi'$, need knowledge of many diagonal and 
off-diagonal amplitudes which are easily summed only if one uses the 
oversimplified oscillator wave functions and a $q\bar q$-proton cross 
section of the form $\sigma_{q\bar q}(\rT)\propto\rT^2$, where $\rT$ is
the transverse $q\bar q$ separation.

Instead, one may switch to the quark basis, which should be equivalent to
the hadronic basis because of completeness. In this representation the
procedure of extracting $\sigma^{\Jpsi\,p}_{tot}$ from photoproduction data
cannot be realized directly, but has to be replaced by a different strategy.
Namely, as soon as one has expressions for the wave functions of charmonia and
the universal dipole cross section $\sigma_{q\bar q}(\rT,s)$, one can predict
both, the experimentally known charmonium photoproduction cross sections and
the unknown $\sigma^{\Jpsi(\psi')\,p}_{tot}$. If the photoproduction data
are well described one may have some confidence in the predictions for the
$\sigma^{\Jpsi(\psi') p}_{tot}$. Of course this procedure will be model
dependent, but we believe that this is the best use of photoproduction
data one can presently make. 
This program was performed for the first time in \cite{KZ}. 
The aim of this paper is not to propose a conceptually new scheme,
but to calculate within a given approach as accurately as possible and
without any free parameters. Wherever there is room for arbitrariness, like
forms for the color dipole cross section and those for for charmonium wave
functions, we use and compare other author's proposals, which have been tested
on data different from those used here.

In the light-cone 
dipole approach the two processes, photoproduction and charmonium-nucleon
elastic scattering look as shown in Fig.~\ref{Fig-D} \cite{KZ}.
\BF
\centerline{
  \scalebox{0.45}{\includegraphics{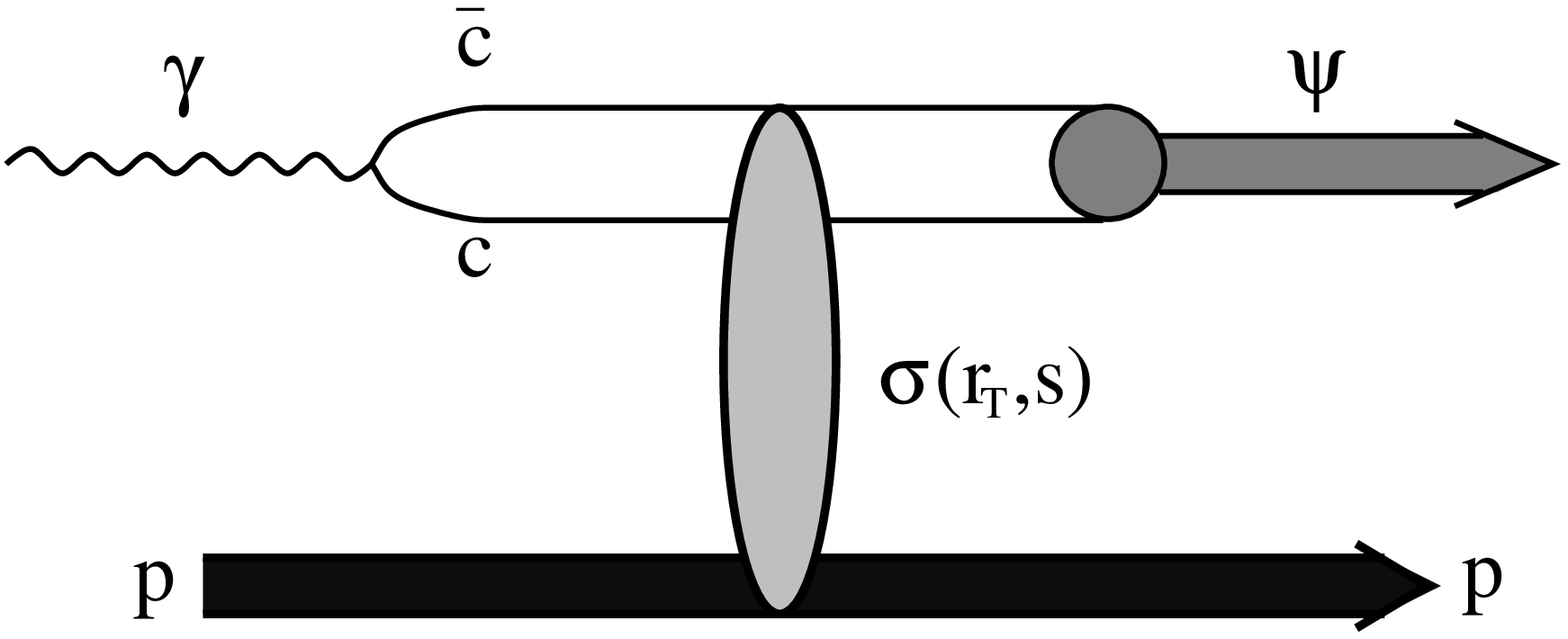}}~~
  \scalebox{0.45}{\includegraphics{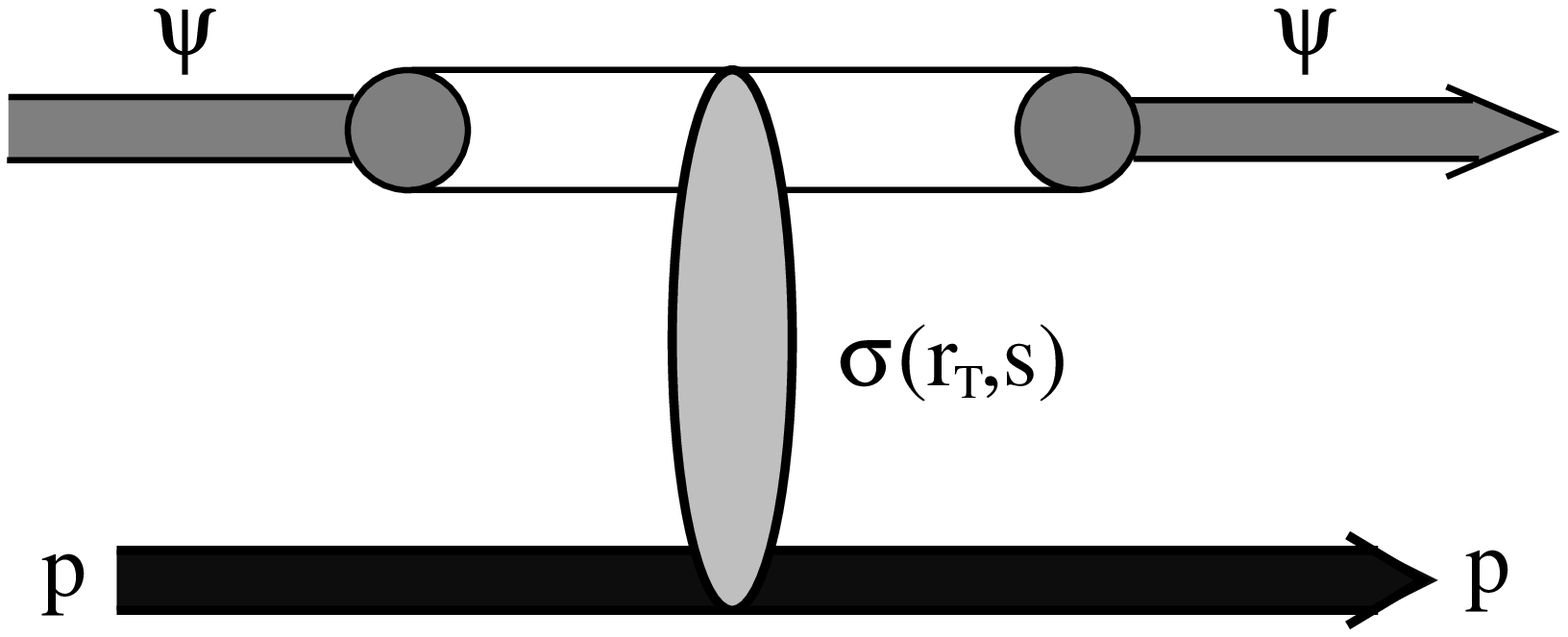}}
}
\Caption{
  \label{Fig-D}
  Schematic representation of the amplitudes for the reactions $\gamma^*p%
  \to \psi p$ (left) and $\psi\,p$ elastic scattering (right) in the rest
  frame of the proton. The $c\bar c$ fluctuation of the photon and the $\psi$
  with transverse separation $\rT$ and c.m. energy $\sqrt{s}$ interact with
  the target proton via the cross section $\sigma(\rT,s)$ and produce
  a $\Jpsi$ or $\psi'$.
}
\EF
The corresponding expressions for the forward amplitudes read 
\BA
  \label{Mgam}
  {\cal M}_{\gamma^* p}(s,Q^2) &=& \sum_{\mu,\bar\mu}
   \,\int\limits_0^1 \!\!d\alpha \int d^2\brT
   \,\Phi^{\!*(\mu,\bar\mu)}_{\psi}(\alpha,\brT) \,\sigma_{q\bar q}(\rT,s)
   \,\Phi^{(\mu,\bar\mu)}_{\gamma^*}(\alpha,\brT,Q^2) \ ,\\
  \label{Mpsi}
  {\cal M}_{\psi\,p}(s)    &=& \sum_{\mu,\bar\mu}
    \,\int\limits_0^1 \!\!d\alpha \int d^2\brT 
    \,\Phi^{\!*(\mu,\bar\mu)}_{\psi}(\alpha,\brT) \,\sigma_{q\bar q}(\rT,s)
    \,\Phi^{(\mu,\bar\mu)}_{\psi}(\alpha,\brT) \ .
\EA
Here the summation runs over spin indexes $\mu$, $\bar\mu$ of the $c$ and
$\bar c$ quarks, $Q^2$ is the photon virtuality, $\Phi_{\gamma^*}(\alpha,%
\rT,Q^2)$ is the light-cone distribution function of the photon for a
$c\bar c$ fluctuation of separation $\rT$ and relative fraction $\alpha$ of
the photon light-cone momentum carried by $c$ or $\bar c$. Correspondingly,
$\Phi_{\psi}(\alpha,\brT)$ is the light-cone wave function of $\Jpsi$,
$\psi'$ and $\chi$ (only in Eq.~\ref{Mpsi}). The dipole cross section
$\sigma_{q\bar q}(\rT,s)$ mediates the transition ({\it cf\/}
Fig.~\ref{Fig-D}).

In Section~\ref{lc-formalism} we review the status of the factorized
light-cone approach to photoproduction of heavy quarkonia. Besides the
well known distribution function of quarks in the photon, it needs
knowledge of the universal flavor independent dipole cross section
which depends on the transverse $\bar qq$ separation and energy.
In Section~\ref{sec-cross} we introduce two parameterizations
available in the literature.

Making use of the nonrelativistic approximation for heavy quarkonia in
Section~\ref{sec-wave} we solve the Schr\"odinger equation with four
types of relativistic potentials available in the literature. The next
most difficult step is a Lorentz boost to the infinite momentum frame
discussed in Section~\ref{sec-spin}. Although this procedure is ill
defined and no unambiguous recipe is known, we apply the standard and
widely used one. We put a special emphasis on importance of the Melosh
spin transformation, which turns out to be very important.

The final expression for the photoproduction cross sections is presented in
Section~\ref{final} and results are compared with available data for $J/\psi$
production in Section~\ref{s-q2}. Although the calculations are parameter
free they demonstrate a very good agreement with data. 

The ratio of $\psi'$ to $J/\psi$ photoproduction yields has drawn attention
recently since previous calculations grossly underestimate experimental values.
It is demonstrated in Section~\ref{ratio} that the Melosh spin transformation
which has been overlooked previously, and accompanies the Lorentz boost may
be the reason. It has a dramatic impact on the $\psi'$ photoproduction
increasing its yield by a factor $2-3$ in a good agreement with data.

After we will have demonstrated that the approach under discussion 
quantitatively explains the photoproduction data, we calculate in 
Section~\ref{psi-n} the total charmonium-nucleon cross sections for
$J/\psi, \psi'$ and $\chi$-s. We predict quite a steep energy dependence
for these cross sections slightly varying for different charmonia.
Although the cross sections correlate with the mean charmonium size,
this dependence is slower than $\propto\la\rT^2\ra$, and this fact
finds a simple explanation. In Section~\ref{nuclei} we compare our
estimates for charmonium-nucleon cross sections with the effective
absorption cross section of charmonium which can be extracted from
data on nuclear attenuation of $J/\psi$ and $\psi'$. Agreement is
rather good.

Our results are summarized in Section~\ref{summary} where we also discuss
the physics of energy dependence of the cross sections and the status of
our approach. Special attention is given to nuclear attenuation of charmonia
which is affected by formation and coherence time phenomena in an important
way.

\section{Light-cone dipole formalism for virtual photoproduction
of charmonia off nucleons}\label{lc-formalism}

The light cone variable describing longitudinal motion which is invariant
to Lorentz boosts is the fraction $\alpha=p_c^+/p_{\gamma^*}^+$ of the
photon light-cone momentum $p_{\gamma^*}^+ = E_{\gamma^*}+p_{\gamma^*}$
carried by the quark or antiquark. In the nonrelativistic approximation
(assuming no relative motion of $c$ and $\bar c$) $\alpha=1/2$ (e.g.
\cite{KZ}), otherwise one should integrate over $\alpha$ (see Eq.~\Ref{Mgam}).
For transversely ($T$) and longitudinally ($L$) polarized photons  
the perturbative photon-quark distribution function in Eq.~\Ref{Mgam} 
reads \cite{ks,bks},
\BE
  \label{psi-g}
  \Phi_{T,L}^{(\mu,\bar\mu)}(\alpha,\brT,Q^2) =
    \frac{\sqrt{N_c\,\alpha_{em}}}{2\,\pi}\,Z_c
    \,\chi_c^{\mu\dagger}\,\widehat O_{T,L}
    \,\widetilde\chi_{\bar c}^{\bar\mu}\,K_0(\epsilon\rT) \ ,
\EE
where 
\BE
  \label{tildechi}
  \widetilde\chi_{\bar c} = i\,\sigma_y\,\chi^*_{\bar c}\ ;
\EE
$\chi$ and $\bar\chi$ are the spinors of the $c$-quark and antiquark
respectively; $Z_c=2/3$. $K_0(\epsilon\rT)$ is the modified Bessel 
function with
\BE
  \label{eps-Q}
  \epsilon^2 = \alpha(1-\alpha)Q^2 + m_c^2\ .
\EE
The operators $\widehat O_{T,L}$ have the form:
\BA
  \widehat O_{T} &=& m_c \, \vec\sigma\cdot\vec e_\gamma
    + i(1-2\alpha)
      \,(\vec\sigma \cdot \vec n)
      \,(\vec e_\gamma \cdot \vec\nabla_{\rT})
    + (\vec n \times \vec e_\gamma)\cdot\vec\nabla_{\rT}\ ,\\
  \widehat O_{L} &=& 2\,Q\,\alpha(1-\alpha)\,\vec\sigma\cdot\vec n\ ,
\EA
where $\vec n=\vec p/p$ is a unit vector parallel to the photon momentum
and $\vec e$ is the polarization vector of the photon. Effects of the
non-perturbative interaction within the $q\bar q$ fluctuation are 
negligible for the heavy charmed quarks.

The color dipole cross section $\sigma_{q\bar q}(\rT,s)$ is poorly known
from first principles. It is expected to vanish $\propto \rT^2$ at small
$\rT\!\!\to\!0$ due to color screening \cite{ZKL} and to level off at
large separations due to a finite range of gluon propagation. We employ 
phenomenological approaches described in section \ref{sec-cross}.

The charmonium wave function is well defined in the rest frame where
one can rely on the Schr\"odinger equation. We present solutions for
four potentials proposed in the literature (section \ref{sec-wave}).
As soon as the rest frame wave function is known, one may be tempted
to apply the Lorentz transformation to the $c\bar c$ pair as it would
be a classical system and boost it to the infinite momentum frame.
However, quantum effects are important and in the infinite momentum
frame a series of different Fock states emerges from the Lorentz boost.
(Compare with a Lorentz boost of a positronium: Weizs\"aker-Williams
photons appear.) Therefore the lowest $|c\bar c\ra$ component in the
infinite momentum frame does not represent the $|c\bar c\ra$ in the
rest frame. We rely on the widely used procedure for the generation
of the light-cone wave functions of charmonia and describe it in 
section \ref{sec-spin}.

\subsection{Phenomenological dipole cross section}
\label{sec-cross}

The dipole formalism for hadronic interactions introduced in \cite{ZKL}
expands the hadronic cross section over the eigen states of the interaction
which in QCD are the dipoles with a definite transverse separation (see
\Ref{Mgam}). Correspondingly, the values of the dipole cross section
$\sigma_{q\bar q}(\rT)$ for different $\rT$ are the eigenvalues of the
elastic amplitude operator. This cross section is flavor invariant, due
to universality of the QCD coupling, and vanishes like $\sigma_{q\bar q}%
(\rT) \propto r^2_T$ for $\rT\!\!\to\!0$. The latter property
is sometimes referred to as color transparency.

The total cross sections for all hadrons and (virtual) photons are known
to rise with energy. Apparently, the energy dependence cannot originate 
from the hadronic wave functions in Eqs.~(\ref{Mgam}, \ref{Mpsi}), but
only from the dipole cross section. In the approximation of two-gluon
exchange used in \cite{ZKL} the dipole cross section is constant, the
energy dependence originates from higher order corrections related to
gluon radiation. On the other way, one can stay with two-gluon exchange,
but involve higher Fock states which contain gluons in addition to the
$q\bar q$. Both approaches correspond to the same set of Feynman graphs.
We prefer to introduce energy dependence into $\sigma_{q\bar q}(\rT,s)$
and not include higher Fock states into the wave functions.

For small size dipoles essential for DIS one may apply perturbative QCD
and the energy dependence comes as an effect of of gluon radiation treated 
in the leading-$\log(1/x)$ approximation \cite{BFKL,book}. In the opposite
limit of large separations typical for light hadrons one can also calculate
the effects of gluon bremsstrahlung making use of smallness of the quark-gluon
correlation radius \cite{k3p}. 

However, the intermediate case we are interested in, is the most complicated
one as usual. No reliable way to sum up higher order corrections is known so
far. Therefore we use a phenomenological form which interpolates between the
two limiting cases of small and large separations. Few parameterizations
are available in the literature, we choose two of them which are simple,
but quite successful in describing data and denote them by the initials of
the authors as ``GBW'' \cite{GBW} and ``KST'' \cite{KST}.

We have
\BA
  \label{GBW}
  \mbox{``GBW'':}~~~~~~~~
  \sigma_{q\bar q}(\rT,x)&=&23.03\left[1-e^{-\rT^2/r_0^2(x)}\right]\mb\ ,\\
  r_0(x) &=& 0.4 \left(\frac{x}{x_0}\right)^{\!\!0.144} \fm\ ,
  \nonumber
\EA
where $x_0=3.04\cdot10^{-4}$. The proton structure function calculated with
this parameterization fits very well all available data at small $x$ and for
a wide range of $Q^2$ \cite{GBW}. However, it obviously fails describing the
hadronic total cross sections, since it never exceeds the value $23.03\mb$.
The $x$-dependence guarantees Bjorken scaling for DIS at high $Q^2$, however,
Bjorken $x$ is not a well defined quantity in the soft limit. Instead we use
the prescription of \cite{Ryskin}, $x=(M^2_\psi+Q^2)/s$, where $M_\psi$ is
the charmonium mass.

This problem as well as the difficulty with the definition of $x$ have been
fixed in \cite{KST}. The dipole cross section is treated as a function
of the c.m. energy $\sqrt{s}$, rather than $x$, since $\sqrt{s}$ is more
appropriate for hadronic processes. A similarly simple form for the dipole
cross section is used
\BA
  \label{KST}
  \mbox{``KST'':}~~~~~~~~~
  \sigma_{\bar qq}(\rT,s) &=& \sigma_0(s) \left[1 - e^{-\rT^2/r_0^2(s)}
  \right]\ .~~~~~~
\EA
The values and energy dependence of hadronic cross sections is guaranteed
by the choice of
\BA
  \sigma_0(s) &=& 23.6 \left(\frac{s}{s_0}\right)^{\!\!0.08} 
  \left(1+\frac38 \frac{r_0^2(s)}{\left<r^2_{ch}\right>}\right)\mb\ ,\\
  r_0(s)      &=& 0.88 \left(\frac{s}{s_0}\right)^{\!\!-0.14}  \fm\ .
\EA
The energy dependent radius $r_0(s)$ is fitted to data for the proton
structure function $F^p_2(x,Q^2)$, $s_0 = 1000\GeV^2$ and the mean square of
the pion charge radius $\left<r^2_{ch}\right>=0.44\fm^2$. The improvement at
large separations leads to a somewhat worse description of the proton structure
function at large $Q^2$. Apparently, the cross section dependent on energy,
rather than $x$, cannot provide Bjorken scaling. Indeed, parameterization
\Ref{KST} is successful only up to $Q^2\approx 10\GeV^2$. 

In fact, the cases we are interested in, charmonium production and
interaction, are just in between the regions where either of these
parameterization is successful. Therefore, we suppose that the difference
between predictions using Eq.~\Ref{GBW} and \Ref{KST} is a measure of the
theoretical uncertainty which fortunately turns out to be rather small.

We demontrate in Fig.~\ref{Fig-dipole} 
%
\BF 
\centerline{\scalebox{0.7}{\includegraphics{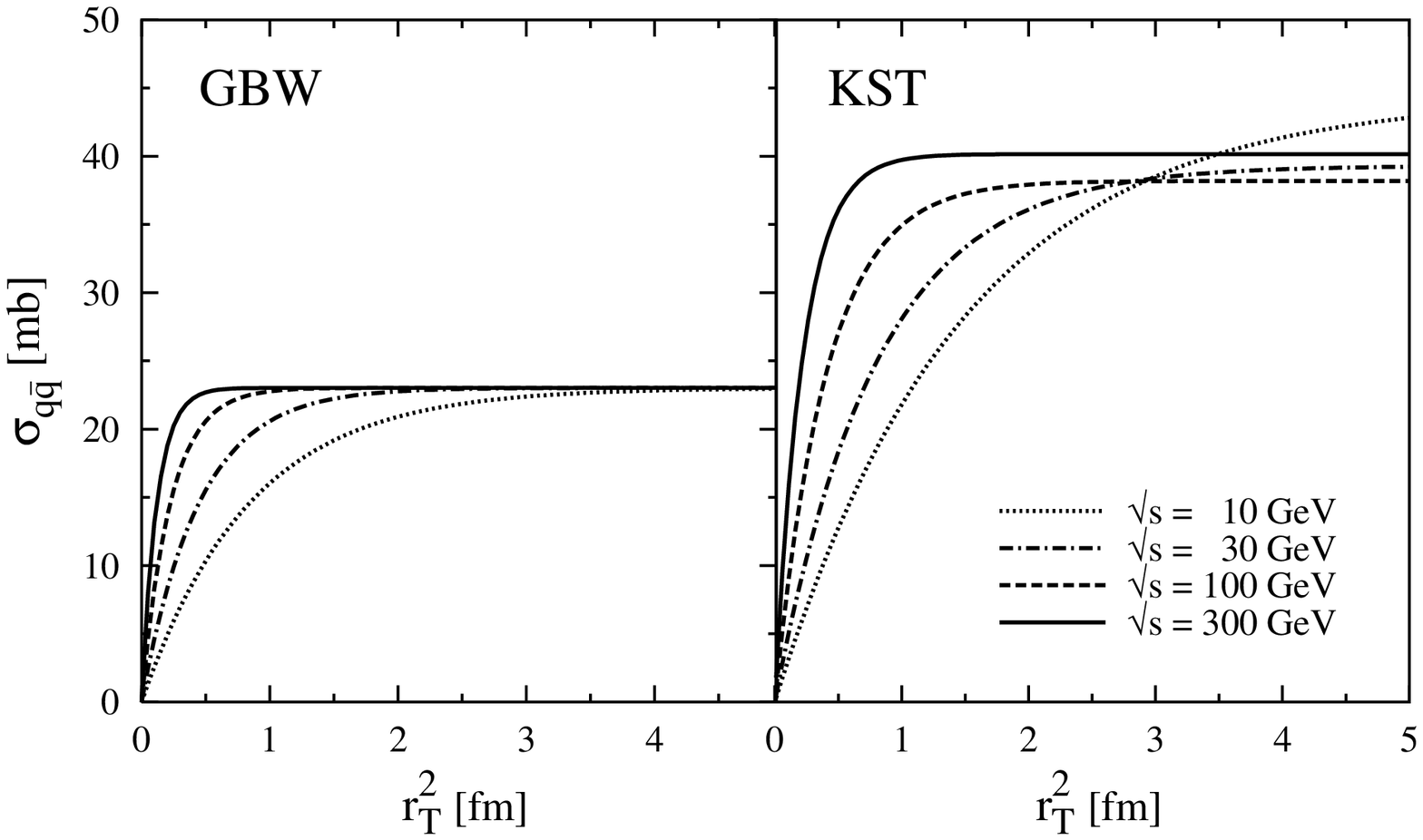}}}
\Caption{
\label{Fig-dipole}
The dipole cross section as function of $r_T^2$ at energies
$\sqrt{s}=10,\ 30,\ 100$ and $300\ \GeV$ for GBW (left) and KST (right) 
parameterizations.
}
\EF
a few examples of $r_T^2$-dependence of the dipole cross section
at different energies for both parameterization. The KST cross section
reveals a nontrivial behavior, it rises with energy at $\rT<3\,fm^2$,
but decreases at larger separations. This is however a temporary
effect, $\sigma_0(s)$ reaches minimum at $\sqrt{s}\approx 77\,GeV$ and
then slowly rises at higher energies. Such a peculiar behavior is a
consequence of our original intention to reproduce the energy dependence of the
hadronic cross sections $\sigma^{hp}_{tot} \propto s^{0.08}$
keeping the form (\ref{KST}) of the cross section. Of course data
are insensitive to the cross section at such large separations.

Both GBW and KST cross section vanish $\propto \rT^2$ at small $\rT$, however
considerably deviate from this simple behavior at large separations.
Quite often, the simplest parameterization ($\propto\rT^2$) for 
the dipole cross section is used. For the coefficient in front of $\rT^2$
we employ the expression obtained by the first term of Taylor expansion of
Eq.~\Ref{KST}:
\BA
  \label{rT2}
  \mbox{``$\rT^2$'':}~~~~~~~~~~~~~
  \sigma_{\bar qq}(\rT,s) &=& \frac{\sigma_0(s)}{r_0^2(s)} \cdot \rT^2 \ .
  ~~~~~~~~~~~~~~~~~~~
\EA

\subsection{Charmonium wave functions}
\label{sec-wave}

The spatial part of the $c\bar c$ pair wave function satisfying the 
Schr\"odinger equation
\BE
  \label{Schroed}
  \left(-\,\frac{\Delta}{m_c}+V(r)\right)
   \,\Psi_{nlm}(\vr)=E_{nl}
   \,\Psi_{nlm}(\vr)
\EE
is represented in the form
\BE
  \label{wf}
  \Psi(\vr) = \Psi_{nl}(r) \cdot Y_{lm}(\theta,\varphi) \ ,
\EE
where $\vr$ is 3-dimensional $c\bar c$ separation, $\Psi_{nl}(r)$ and
$Y_{lm}(\theta,\varphi)$ are the radial and orbital parts of the wave
function. The equation for radial $\Psi(r)$ is solved with the help
of the program \cite{Lucha}. The following four potentials $V(r)$ 
have been used (see Fig.~\ref{Fig-Vr}):

\begin{itemize}
\item ``COR'': Cornell potential \cite{COR},
  \BE
    \label{COR}
    V(r) = -\frac{k}{r} + \frac{r}{a^2}
  \EE
  with $k=0.52$, $a=2.34\GeV^{-1}$ and $m_c=1.84\GeV$.
\item ``BT'': Potential suggested by Buchm\"uller and Tye \cite{BT}
  with $m_c=1.48\GeV$. It has a similar structure as the Cornell
  potential: linear string potential at large separations and 
  Coulomb shape at short distances with some refinements, however.
\item ``LOG'': Logarithmic potential \cite{LOG}
  \BE
    \label{LOG}
    V(r) = -0.6635\GeV + (0.733\GeV) \log(r \cdot 1\GeV)
  \EE
  with $m_c=1.5\GeV$.
\item ``POW'': Power-law potential \cite{POW}
  \BE
    \label{POW}
    V(r) = -8.064\GeV + (6.898\GeV) (r \cdot 1\GeV)^{0.1}
  \EE
  with $m_c=1.8\GeV$.
\end{itemize}
The shapes of the four potentials is displayed in Fig.~\ref{Fig-Vr} and
differ from each other only at large $r$ ($\geq 1\fm$) and very small
$r$ ($\leq 0.05\fm$) separations. Note, however, that COR and POW use 
$m_c \approx 1.8\GeV$, while BT and LOG use $m_c \approx 1.5\GeV$
for the mass of the charmed quark. This difference will have
significant consequences.

\BF
\centerline{\scalebox{0.7}{\includegraphics{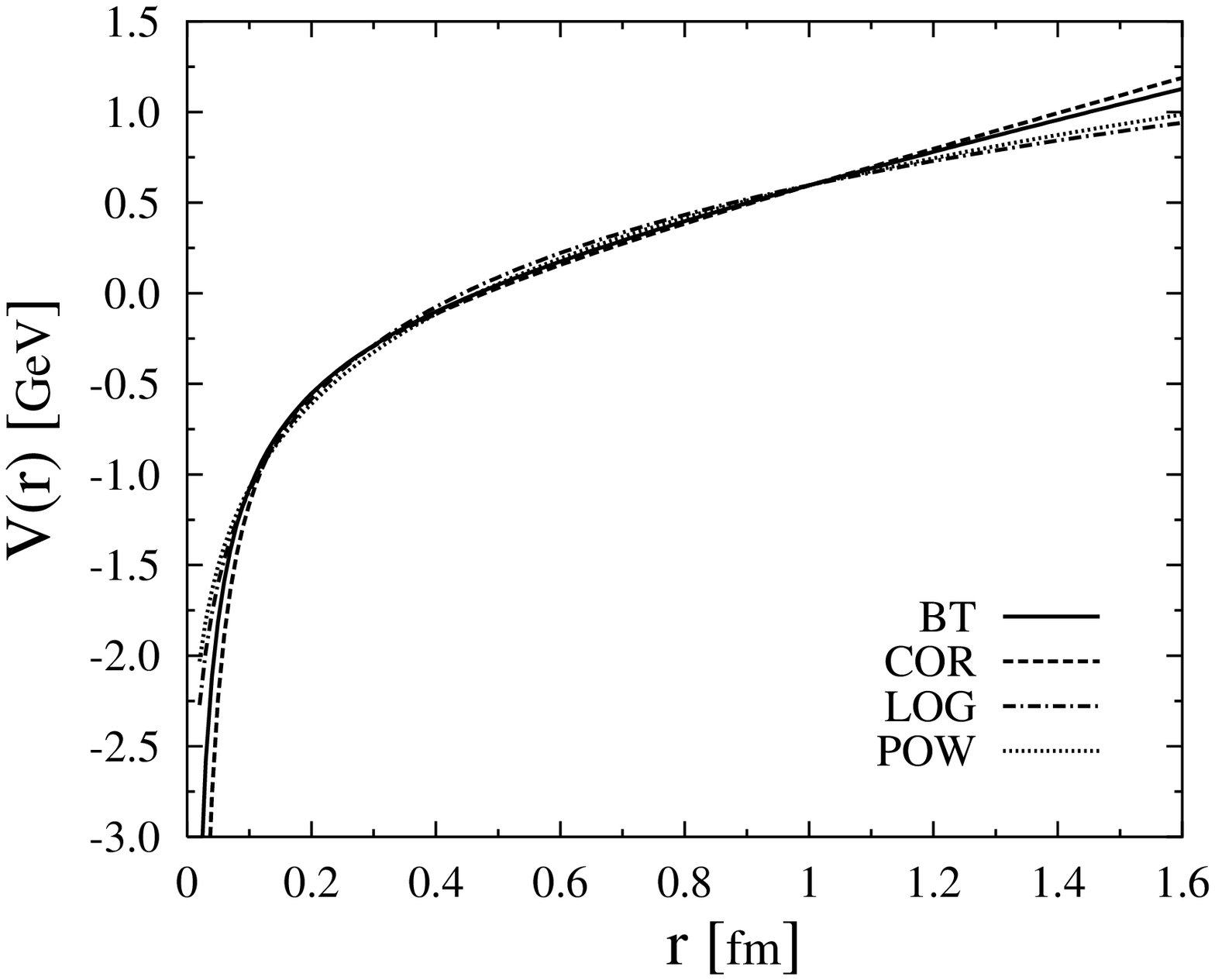}}}
\Caption{
  \label{Fig-Vr}
  Shapes of the potentials $V(r)$ for the four parameterizations
  employed in this paper. The curves for COR, LOG and POW are
  normalized at $r=1\fm$ to the value of BT potential.
}
\EF

The results of calculations for the radial part $\Psi_{nl}(r)$ of the $1S$
and $2S$ states are depicted in Fig.~\ref{Fig-y}. For the ground state all
the potentials provide a very similar behavior for $r>0.3\fm$, while for 
small $r$ the predictions are differ by up to $30\%$. The peculiar property
of the $2S$ state wave function is the node at $r\approx 0.4\fm$ which causes
strong cancelations in the matrix elements Eq.~\Ref{Mgam} and as a result,
a suppression of photoproduction of $\psi'$ relative to $\Jpsi$
\cite{KZ,Benhar}.

\BF
\centerline{
  \scalebox{0.47}{\includegraphics{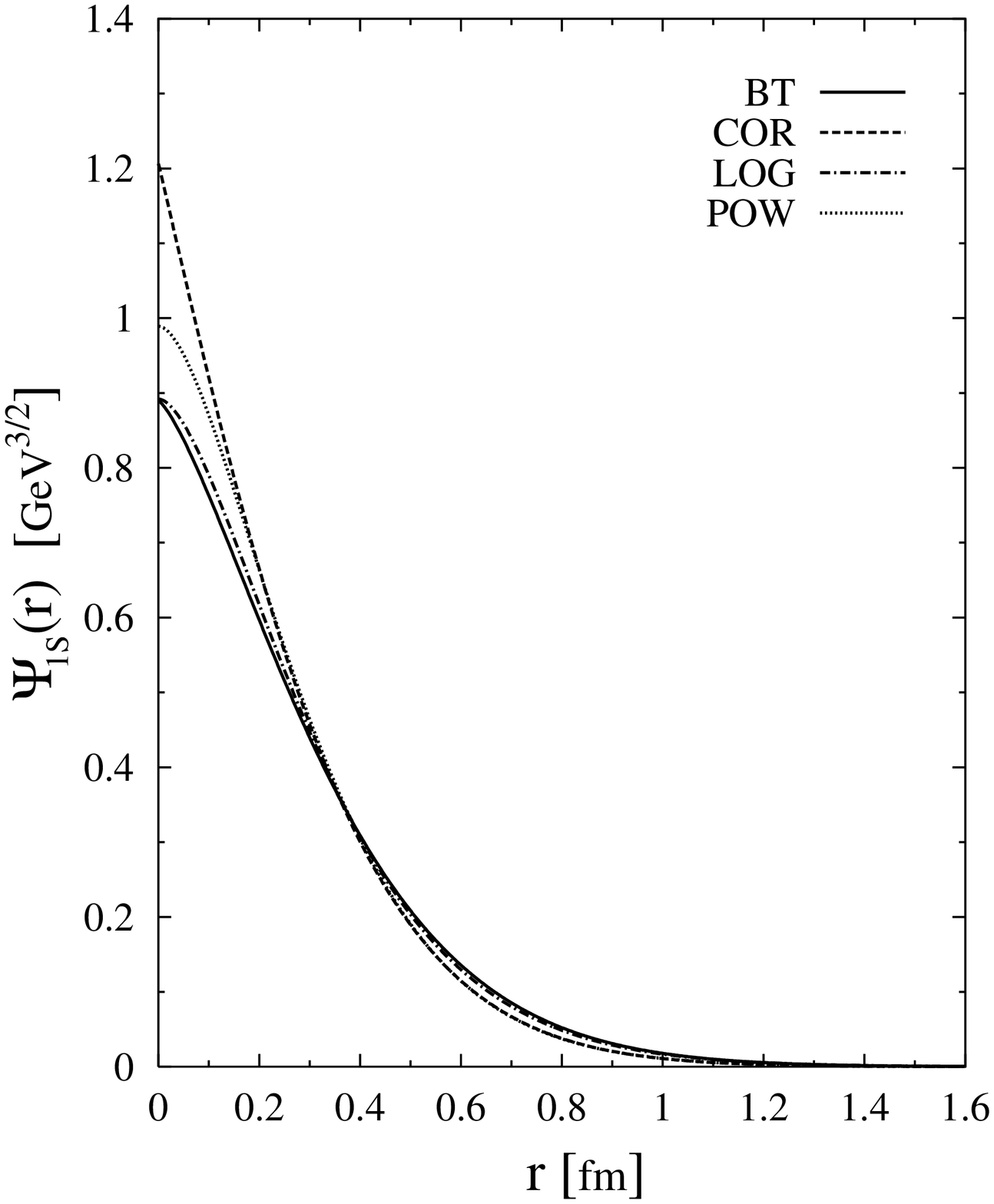}}~~
  \scalebox{0.47}{\includegraphics{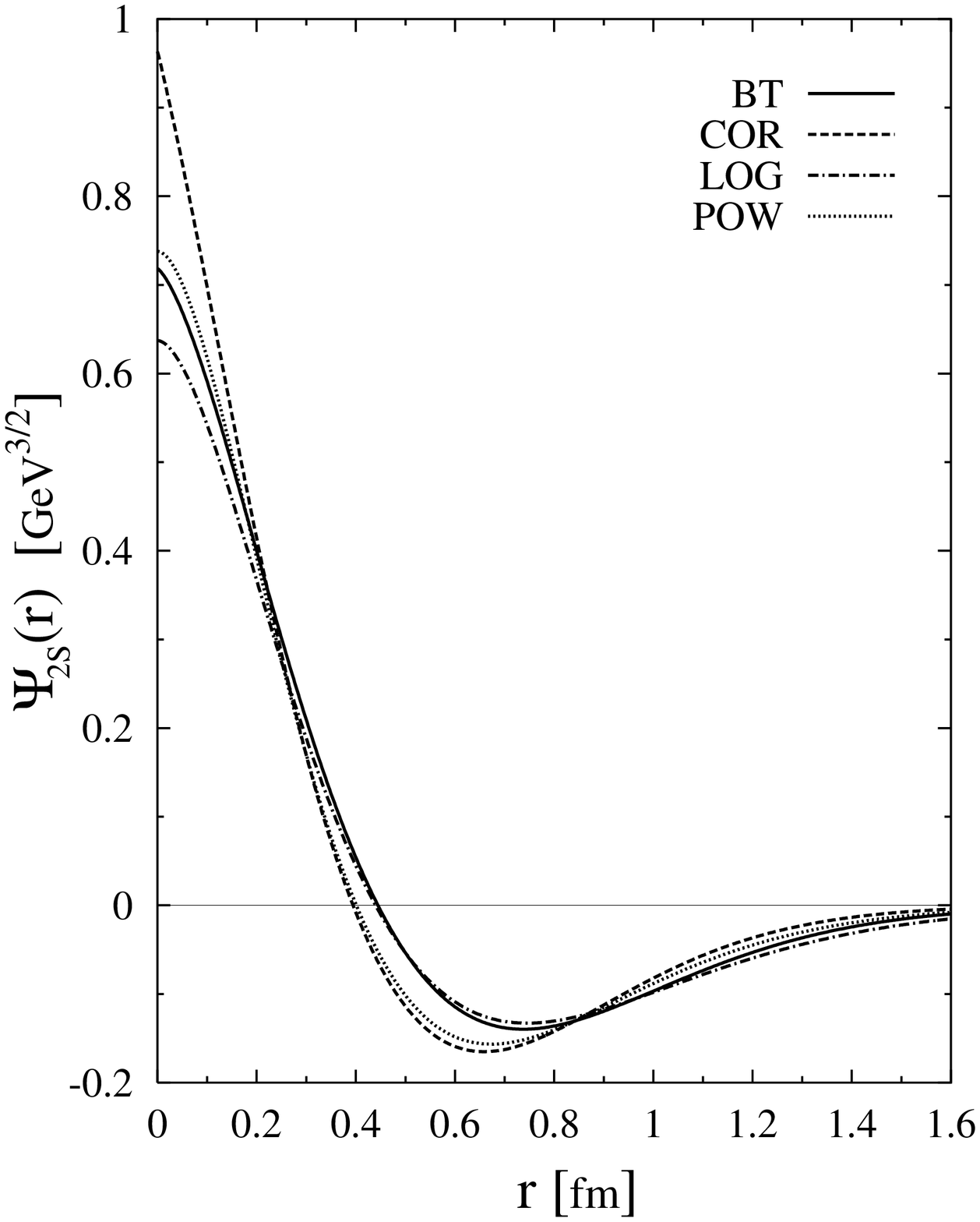}}
}
\Caption{
\label{Fig-y}
  The radial part of the wave function $\Psi_{nl}(r)$ for the $1S$ and
  $2S$ states calculated with four different potentials (see text).
}
\EF

\subsection{Light-cone wave functions for the bound states}\label{sec-spin}

As has been mentioned, the lowest Fock component $|c\bar c\ra$ in the
infinite momentum frame is not related by simple Lorentz boost to the
wave function of charmonium in the rest frame. This makes the problem
of building the light-cone wave function for the lowest $|c\bar c\ra$
component difficult, no unambiguous solution is yet known. There are
only recipes in the literature, a simple one widely used \cite{Terent'ev},
is the following. One applies a Fourier transformation from coordinate to
momentum space to the known spatial part of the non-relativistic wave
function \Ref{wf}, $\Psi(\vr)\Rightarrow\Psi(\vp)$, which can be written
as a function of the effective mass of the $c\bar c$, $M^2=4(p^2+m_c^2)$,
expressed in terms of light-cone variables
\BE
  M^2(\alpha,\pT) = \frac{\pT^2+m_c^2}{\alpha(1-\alpha)}\ .
\EE 
In order to change integration variable $p_L$ to the light-cone variable
$\alpha$ one relates them via $M$, namely $p_L=(\alpha-1/2)M(p_T,\alpha)$.
In this way the $c\bar c$ wave function acquires a kinematical factor
\BE
  \label{lc-wf-p}
  \Psi(\vp) \Rightarrow
  \sqrt{2}\,\frac{(p^2+m_c^2)^{3/4}}{(\pT^2+m_c^2)^{1/2}}
  \cdot \Psi(\alpha,\bpT)
  \equiv \Phi_\psi(\alpha,\bpT) \ .
\EE

This procedure is used in \cite{Hoyer} and the result is applied to 
calculation of the amplitudes \Ref{Mgam}. The result is discouraging,
since the $\psi'$ to $\Jpsi$ ratio of the photoproduction cross sections
are far too low in comparison with data. However, the oversimplified
dipole cross section $\sigma_{q\bar q}(\rT)\propto\rT^2$ has been used,
and what is even more essential, the important ingredient of Lorentz
transformations, the Melosh spin rotation, has been left out. The spin
transformation has also been left out in the recent publication \cite{Suzuki}
which repeats the calculations of \cite{Hoyer} with a more realistic
dipole cross section which levels off at large separations. This leads
to suppression of the node-effect (less cancelation) and enhancement
of $\Psi'$ photoproduction. Nevertheless, the calculated $\psi'$ to
$\Jpsi$ ratio is smaller than the data by a factor of two.

The 2-dimensional spinors $\chi_c$ and $\chi_{\bar c}$ describing $c$
and $\bar c$ respectively in the infinite momentum frame are known to be
related via the Melosh rotation \cite{Melosh,Terent'ev} to the spinors 
$\bar\chi_c$ and $\bar\chi_{\bar c}$ in the rest frame:
\BA
  \nonumber
  \bf\overline{\chi}_c        &=& \widehat R(  \alpha, \bpT)\,\chi_c\ ,\\
  \bf\overline{\chi}_{\bar c} &=& \widehat R(1-\alpha,-\bpT)\,\chi_{\bar c}\ ,
  \label{Melosh}
\EA
where the matrix $R(\alpha,\bpT)$ has the form:
\BE
  \widehat R(\alpha,\bpT) = 
    \frac{  m_c+\alpha\,M - i\,[\vec\sigma \times \vec n]\,\bpT}
    {\sqrt{(m_c+\alpha\,M)^2+\pT^2}} \ .
\label{matrix}
\EE

Since the potentials we use in section~\ref{sec-wave} contain no spin-orbit
term, the $c\bar c$ pair is in $S$-wave. In this case spatial and spin 
dependences in the wave function factorize and we arrive at the following
light cone wave function of the $c\bar c$ in the infinite momentum frame
\BE
  \label{lc-wf}
  \Phi^{(\mu,\bar\mu)}_\psi(\alpha,\bpT) =
     U^{(\mu,\bar\mu)}(\alpha,\bpT)\cdot\Phi_\psi(\alpha,\bpT)\ ,
\EE
where 
\BE
  U^{(\mu,\bar\mu)}(\alpha,\bpT) = 
    \chi_{c}^{\mu\dagger}\,\widehat R^{\dagger}(\alpha,\bpT)
    \,\vec\sigma\cdot\vec e_\psi\,\sigma_y
    \,\widehat R^*(1-\alpha,-\bpT)
    \,\sigma_y^{-1}\,\widetilde\chi_{\bar c}^{\bar\mu}
\EE
and $\widetilde\chi_{\bar c}$ is defined in \Ref{tildechi}.

Note that the wave function \Ref{lc-wf} is different from one used in
\cite{Ryskin93,fs,Nemchik} where it was assumed that the vertex
$\psi\to c\bar c$ has the structure $\psi_{\mu}\,\bar u\,\gamma_{\mu}\,u$
like the for the photon $\gamma^*\to c\bar c$. The rest frame wave function
corresponding to such a vertex contains $S$ wave and $D$ wave. The weight
of the latter is dictated by the structure of the vertex and cannot be
justified by any reasonable nonrelativistic potential model for the
$c\bar c$ interaction.

Now we can determine the light-cone wave function in the mixed
longitudinal momentum - transverse coordinate representation:
\BE
  \label{lc-wf-r}
  \Phi^{(\mu,\bar\mu)}_\psi(\alpha,\brT) =
    \frac{1}{2\,\pi} 
    \int d^2\bpT\,e^{-i\bpT\brT}\,
    \Phi^{(\mu,\bar\mu)}_\psi(\alpha,\bpT)\ .
\EE

The spatial component $\Phi_\psi(\alpha,\brT)$ of Eq.~\Ref{lc-wf-p} in mixed
representation \Ref{lc-wf-r} is plotted as a function of $\rT$ and $\alpha$
in Fig.~\ref{Fig-f} for $\Jpsi(1S)$ and $\psi'(2S)$ states. While the $1S$
wave function depends monotonically on $\rT$ and smoothly vanishes at small
$\alpha$, the wave function of the $2S$ state demonstrates a nontrivial
behavior: the node disappears for small $\alpha$.
\BF
\centerline{
  \scalebox{0.52}{\includegraphics{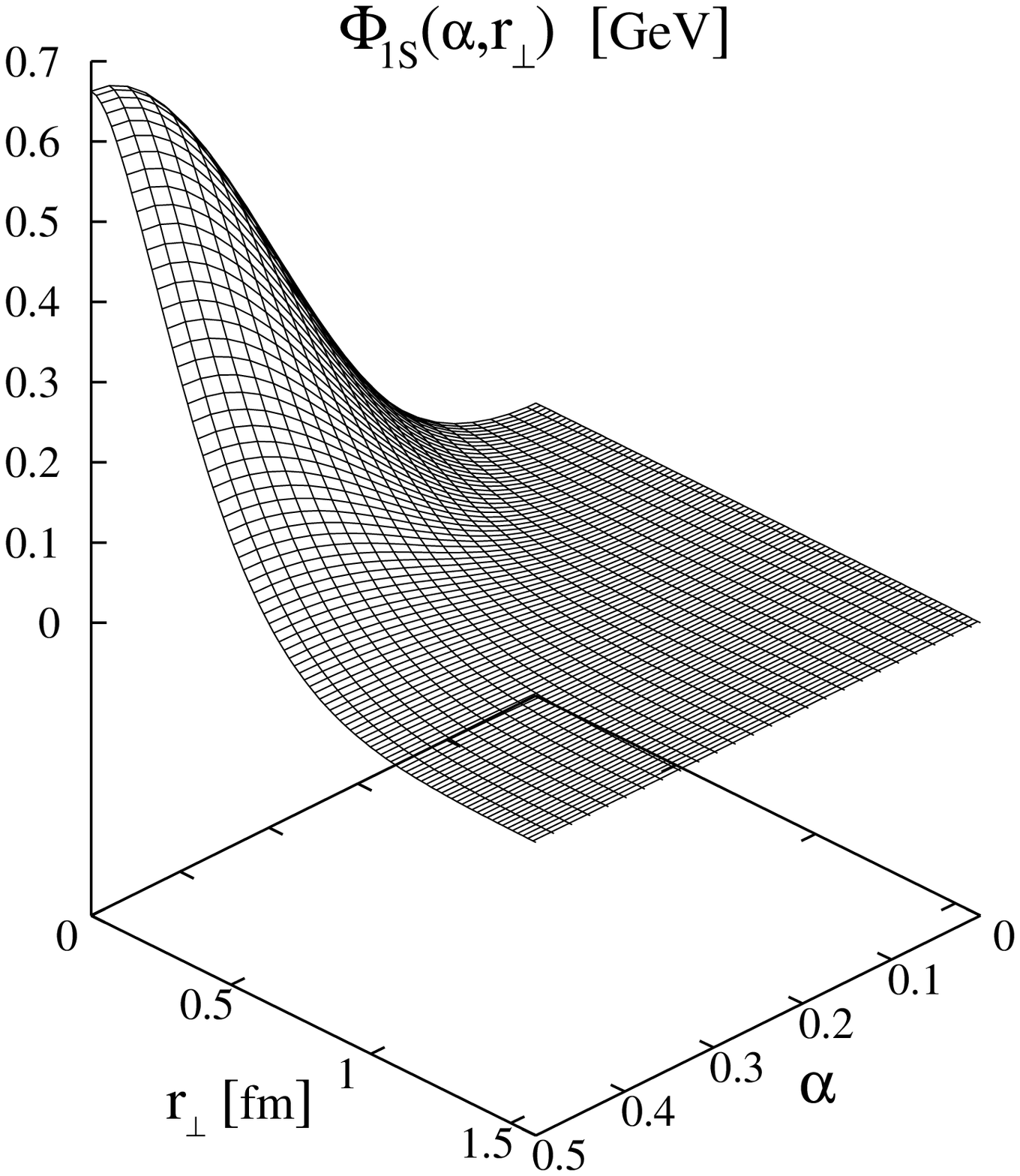}}~~
  \scalebox{0.52}{\includegraphics{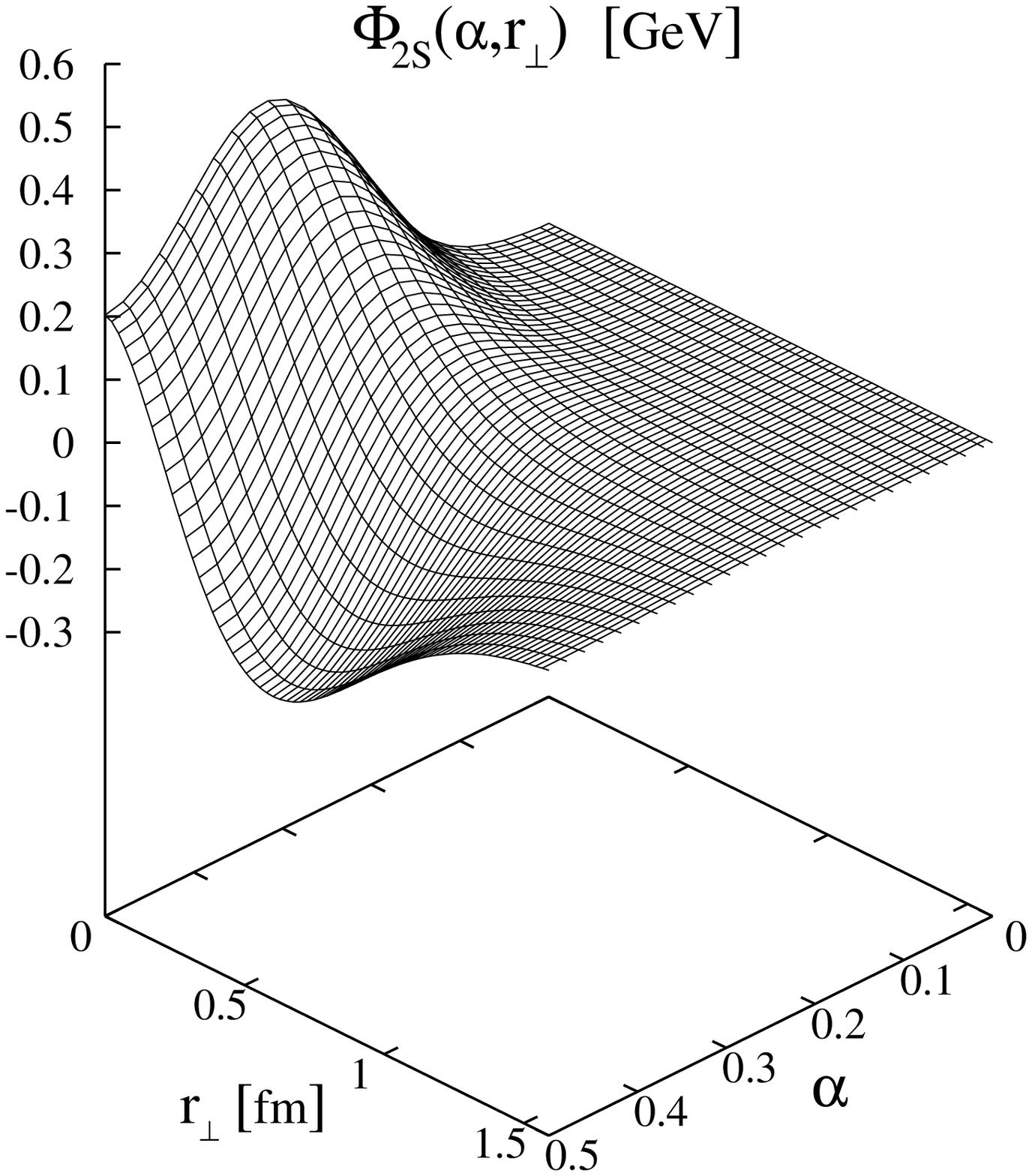}}
}
\Caption{
  \label{Fig-f}
  Three-dimensional plot for the light-cone wave functions for
  $J/\psi(1S)$ and $\psi'(2S)$ in the mixed $\alpha-\br_T$ 
  representation for the BT potential~\cite{BT}.
}
\EF

\section{Calculations and comparison with data}\label{data}

\subsection{The final expressions}\label{final}

Having the light-cone wave function of charmonium in momentum
representation Eq.~\Ref{lc-wf} it is more convenient to switch
to an integration over $\bpT$ in the matrix element Eq.~\Ref{Mgam}:
\BE
  \label{p-repr}
  {\cal M}_{T,L}(s,Q^2) =
    \int\limits_0^1 \!\!d\alpha \int d^2\bpT\,
    \Phi^*_\psi(\alpha,\bpT)\,\Sigma_{T,L}(\alpha,\bpT,s,Q^2)\ ,
\EE
where
\BE
  \Sigma_{T,L}(\alpha,\bpT,s,Q^2) =
    \frac{1}{2\pi} \sum_{\mu,\bar\mu}
    U^{(\mu,\bar\mu)}(\alpha,\bpT)
    \,\int d^2\brT\, e^{i\bpT\brT}
    \,\sigma(\rT,s)
    \,\Phi^{(\mu,\bar\mu)}_{T,L}(\alpha,\brT)\ .
\EE
If the dipole cross section depends on $\rT$ like $\sigma_0\!%
\left(1-e^{-\rT^2/r_0^2}\right)$ (see Eqs.~\Ref{GBW} and \Ref{KST}), then
$\Sigma(\alpha,\bpT,s,Q^2)$ which includes the effects of spin rotation,
can be expressed as follows:
\BA
  \Sigma_T(\alpha,\bpT,s,Q^2) &=&
    \frac{1}{m_c} \left[m_T-\frac{2\,\pT^2\,\alpha(1-\alpha)}{m_T+m_L}\right]\,
    \widetilde\Sigma_T(\alpha,\bpT,s,Q^2)\nonumber \\
    &-&
    \frac{2\,\pT^2}{m_q\, m_T\, r_0^2}
    \left[1+\frac{m_T\, (1-2\alpha)}{m_T+m_L}\right]
    \frac{\partial\widetilde\Sigma_t(\alpha,\bpT,s,Q^2)}
    {\partial\pT^2} \ ,\\
  \Sigma_L(\alpha,\bpT,s,Q^2) &=&
    \frac{m_q^2 + m_T m_L}{m_c (m_T+m_L)}\,
    \widetilde\Sigma_L(\alpha,\bpT,s,Q^2)\ , 
\EA
where $m_T^2=m_c^2+\pT^2$, $m_L^2=4\,m_c^2\,\alpha(1-\alpha)$ and
\BA
  \widetilde\Sigma_{T,L}(\alpha,\bpT,s,Q^2) &=&
    \frac{\sigma_0(s)}{2\pi} \int d^2\rT\, e^{i\bpT\brT}
    \,\Phi_{T,L}(\alpha,\brT,Q^2)
    \cdot \left[1 - e^{-\rT^2/r_0^2(s)}\right],\\
  \widetilde\Sigma_t(\alpha,\bpT,s,Q^2) &=&
    \frac{\sigma_0(s)}{2\pi} \int d^2\rT\, e^{i\bpT\brT}
    \,\Phi_T(\alpha,\brT,Q^2)
    \cdot e^{-\rT^2/r_0^2(s)}\ ,\\
  \Phi_T(\alpha,\rT,Q^2) &=& 
    \frac1\pi \sqrt\frac{2\alpha_{em}}{3}\,m_q\,K_0(\epsilon\rT)\ ,\\
  \Phi_L(\alpha,\rT,Q^2) &=&
    \frac2\pi \sqrt\frac{ \alpha_{em}}{3}\,Q
                          \,\alpha (1-\alpha) \,K_0(\epsilon\rT)\ .
\EA
 
The photoproduction cross section is given by
\BE
  \label{sigma-gp}
  \sigma_{\gamma^*p\to\psi p}(s,Q^2) = 
  \frac{\vert\widetilde{\cal M}_T(s,Q^2)\vert^2 + 
  \eps\,\vert\widetilde{\cal M}_L(s,Q^2)\vert^2}{16\,\pi\,B}\ ,
\EE
where $\eps$ is the photon polarization (for H1 data $\la\eps\ra=0.99$); 
$B$ is the slope parameter in reaction $\gamma^* p\to\psi p$. We use the
experimental value \cite{H1-s} $B=4.73\,GeV^{-2}$. $\widetilde{\cal M}_{T,L}$
includes also the correction for the real part of the amplitude:
\BE
  \label{re/im}
  \widetilde{\cal M}_{T,L}(s,Q^2) = {\cal M}_{T,L}(s,Q^2)
   \,\left(1 - i\,\frac{\pi}{2}\,\frac{\partial
   \,\ln{\cal M}_{T,L}(s,Q^2)}{\partial\,\ln s} \right)\ ,
\EE
where we apply the well known derivative analyticity relation between the
real and imaginary parts of the forward elastic amplitude \cite{Bronzan}.
The correction from the real part is not small since the cross section of
charmonium photoproduction is a rather steep function of energy (see below).

\subsection{$s$ and $Q^2$ dependence of $\sigma(\gamma^*p \to \Jpsi\,p$)}
\label{s-q2}

Now we are in a position now to calculate the cross section of charmonium
photoproduction using Eq.~(\ref{sigma-gp}).
The results for $J/\psi$ are compared with the data in
Fig.~\ref{Fig-s}. Calculations are performed with GBW and KST parameterizations
for the dipole cross section and for wave functions of the $\Jpsi$ calculated
from BT, LOG, COR and POW potentials. One observes
\begin{itemize}
\item There are no major differences for the results using the GBW and
      KST parameterizations.
\item The use of different potentials to generate the wave functions
      of the $\Jpsi$ leads to two distinctly different behaviors. The
      potentials labeled BT and LOG (see sect. \ref{sec-wave}) describe
      the data very well, while the potentials COR and LOG underestimate
      them by a factor of two. The different behavior has been traced
      to the following origin: BT and LOG use $m_c \approx 1.5\GeV$,
      but COR and POW $m_c \approx 1.8\GeV$. While the bound state
      wave functions of $\Jpsi$ are little affected by this difference
      (see Fig.~\ref{Fig-y}), the photon wave function Eq.~\Ref{psi-g}
      depends sensitively on $m_c$ via the argument Eq.~\Ref{eps-Q} of
      the $K_0$ function.
\end{itemize}
\BF
\centerline{\scalebox{0.92}{\includegraphics{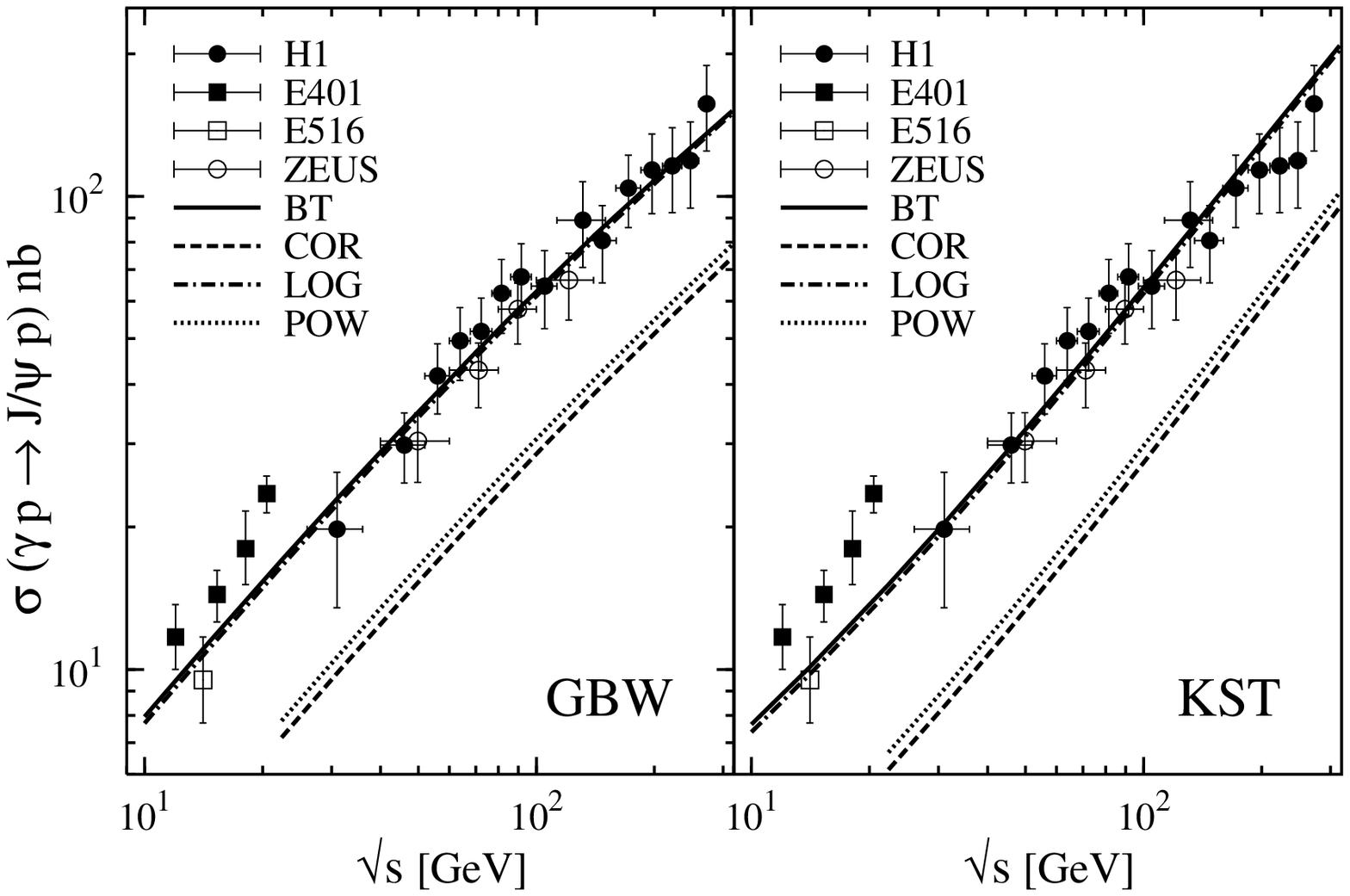}}}
\Caption{
  \label{Fig-s}
  Integrated cross section for elastic photoproduction $\gamma\,p%
  \rightarrow J/\psi\,p$ with real photons ($Q^2=0$) as a function
  of the energy calculated with GBW and KST dipole cross sections
  and for four potentials to generate $\Jpsi$ wave functions.
  Experimental data points from the H1~\cite{H1-s}, E401~\cite{E401-s},
  E516~\cite{E516-s} and ZEUS~\cite{ZEUS-s} experiments.
}
\EF

We compare our calculations also with data for the $Q^2$ dependence
of the cross section. The data are plotted in Fig.~\ref{Fig-QM} at c.m. energy
$\sqrt s=90\GeV$ as a function of $Q^2+M^2_{\Jpsi}$, since in this form both, data 
and calculations display an approximate power law dependence. 
\BF
\centerline{\scalebox{0.8}{\includegraphics{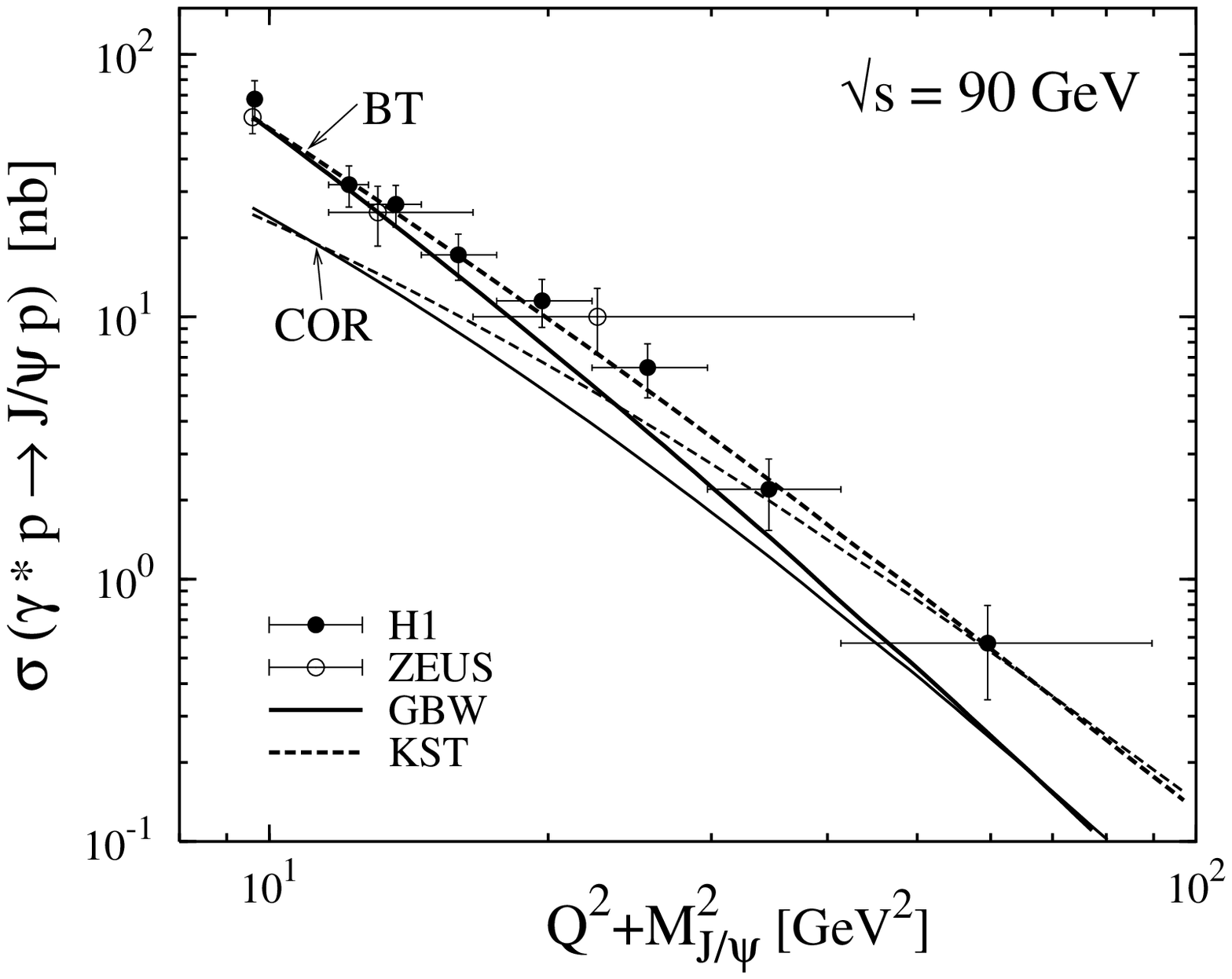}}}
\Caption{
  \label{Fig-QM}
  Integrated cross section for elastic photo production as a function
  of the photon virtuality $Q^2+M_{J/\psi}$ at energy $\sqrt{s}=90\GeV$.
  Solid and dashed curves are calculated with GBW and KST dipole cross
  sections, while thick and thin curves correspond to BT and COR potentials,
  respectively. Results obtained with LOG and POW potentials are very
  close to that curves (LOG similar to BT and POW to COR, see also
  Fig.~\ref{Fig-s}). 
}
\EF
Such a dependence on $Q^2+M^2_{\Jpsi}$ is suggested by the variable
$\epsilon^2$ in Eq.~(\ref{eps-Q}), which for $\alpha=1/2$ takes the value
$Q^2+(2\,m_c)^2$. It may be considered as an indication that $\alpha=1/2$
is a reasonable approximation for the nonrelativistic charmonium wave function.

Our results are depicted for BT and COR potentials and using GBW and KST
cross sections. Agreement with the calculations based on BT potential
is again quite good, while the COR potential grossly underestimate the
data at small $Q^2$. Although the GBW and KST dipole cross sections
lead to nearly the same cross sections for real photoproduction,
their predictions at high $Q^2$ are different by a factor $2-3$.
Supposedly the GBW parameterization should be more trustable
at $Q^2\gg M_{\Psi}^2$.

\subsection{Importance of spin effects for the $\psi'$ to $\Jpsi$
ratio}\label{ratio}

It turns out that the effects
of spin rotation have a gross impact on the cross section of elastic
photoproduction $\gamma\,p \to \Jpsi(\psi)p\,$. To demonstrate these
effects we present the results of our calculations at
$\sqrt{s}=90\,GeV$ in Table~\ref{Tab-sR}.
\BT
\Caption{
  \label{Tab-sR}
  The photoproduction $\gamma\,p \to \Jpsi\,p\,$ cross-section
  $\sigma(\Jpsi)$ in nb and the ratio $R=\sigma(\psi')/\sigma(\Jpsi)$
  for the four different types of potentials (BT, LOG, COR, POW) and
  the three parameterizations (GBW, KST, $\rT^2$) for the dipole cross
  section $\sigma(\rT,s)$ at $\sqrt{s}=90\GeV$. The values in parentheses
  correspond to the case when the spin rotation is neglected.
}
\vskip3mm
\begin{center}
\begin{tabular}{|ll|c|c|c|c|}
\hline
\vphantom{\Bigg\vert}
  &
  & BT
  & LOG
  & COR
  & POW \\
\hline &&&&&\\[-3mm]
\framebox{$\sigma$} 
&
  GBW     & 52.01~(37.77) & 50.78~(36.63) & 23.13~(17.07) & 24.94~(18.64)\\[0.5mm]
& KST     & 49.96~(35.87) & 48.49~(34.57) & 21.05~(15.42) & 22.83~(16.92)\\[0.5mm]
& $\rT^2$ & 66.67~(47.00) & 64.07~(44.86) & 25.81~(18.71) & 28.23~(20.66)\\[1.5mm]
\hline &&&&&\\[-3mm]
\framebox{$R$} &
  GBW     & 0.147~(0.075) & 0.117~(0.060) & 0.168~(0.099) & 0.144~(0.085)\\[0.5mm]
& KST     & 0.147~(0.068) & 0.118~(0.054) & 0.178~(0.099) & 0.152~(0.084)\\[0.5mm]
& $\rT^2$ & 0.101~(0.034) & 0.081~(0.027) & 0.144~(0.070) & 0.121~(0.058)\\[1.5mm]
\hline
\end{tabular}
\end{center}
\ET
The upper half of the table shows the photoproduction cross sections for
$\Jpsi$ for different parameterizations of the dipole cross section (GBW,
KST, ``$\rT^2$'') and potentials (BT, COR, LOG, POW). The numbers in
parenthesis show what the cross section would be, if the spin rotation
effects were neglected. We see that these effects add 30-40\% to the
$\Jpsi$ photoproduction cross section. 

The spin rotation effects turn out to have a much more dramatic impact on
$\psi'$ increasing the photoproduction cross section by a factor 2-3. This
is visible in the lower half of the table which shows the ratio $R=\sigma%
(\psi')/\sigma(\Jpsi)$ of photoproduction cross sections, where the number
in parenthesis correspond to no spin rotation effects included. This spin
effects explain the large values of the ratio $R$ observed experimentally.
Our results for $R$ are about twice as large as evaluated in \cite{Suzuki}
and even more than in \cite{Hoyer}.

The ratio of $\psi'$ to $\Jpsi$ photoproduction cross sections is
depicted as function
of c.m. energy in Fig.~\ref{Fig-Rs} 
and as a function of $Q^2$ in Fig.~\ref{Fig-RQ}
for all four potentials and for the parameterizations of the 
dipole cross sections GBW and KST.
\BF
\centerline{\scalebox{0.88}{\includegraphics{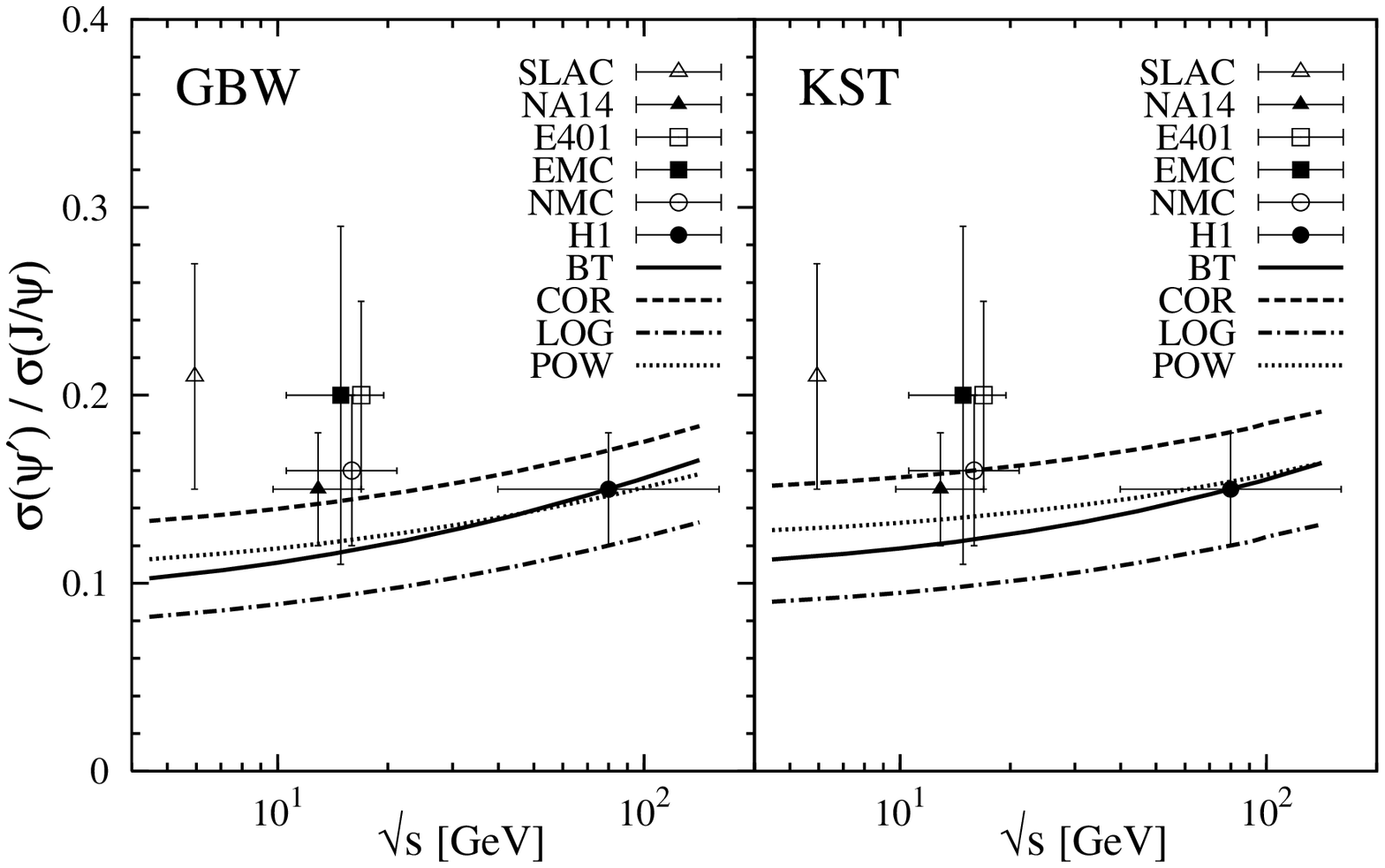}}}
\Caption{
\label{Fig-Rs}
  The ratio of $\psi'$ to $J/\psi$ photoproduction cross sections
  as a function of c.m. energy calculated for all four potentials
  with with GBW and KST parameterizations for the dipole cross
  section. Experimental data points from the SLAC~\cite{SLAC-R},
  NA14~\cite{NA14-R}, E401~\cite{E401-R}, EMC~\cite{EMC-R},
  NMC~\cite{NMC-R} and H1~\cite{H1-R} experiments.
}
\EF

Our calculations agree with available data, but error bars
are too large to provide a more precise test for the theory. Remarkably,
the ratio $R(s)$ rises with energy. This result is in variance with the naive
expectation based on the larger size of the $\psi'$ and on the usual rule:
the smaller the size of the $q\bar q$ dipole, the steeper its energy
dependence. There is, however, no contradiction, since this is another
manifestation of the node in the wave function of $\psi'$. 
Indeed, as function of energy mostly the
short distance part of the dipole cross section $\sigma_{q\bar q}(\rT)$
rises. It enhances the positive contribution for distances
shorter than the node position in the $\psi'$ wave function. Therefore,
with increasing  energy the cancelation in the amplitude
of $\psi'$ production is reduced. This effect leads to a steeper energy dependence
of $\psi'$ production compared to $\Jpsi$. The effect is stronger for GBW
than KST parameterizations, since the GBW cross section does not rise with
energy at all at large separations. Note that this situation is specific
for photoproduction because the nodeless wave function of the photon is
projected to the sign changing wave function of $\psi'$. This should not
happen in the case of elastic $\Jpsi(\Psi')$-$p$ scattering (see below).

Similarly of the node effect leads to a rising $Q^2$-dependence of 
the $\psi'$ to $\Jpsi$ ratio in the photoproduction cross sections. Our
calculations are compared with available data in Fig.~\ref{Fig-RQ} for
the GBW and KST parameterizations respectively.
\BF
\centerline{\scalebox{0.88}{\includegraphics{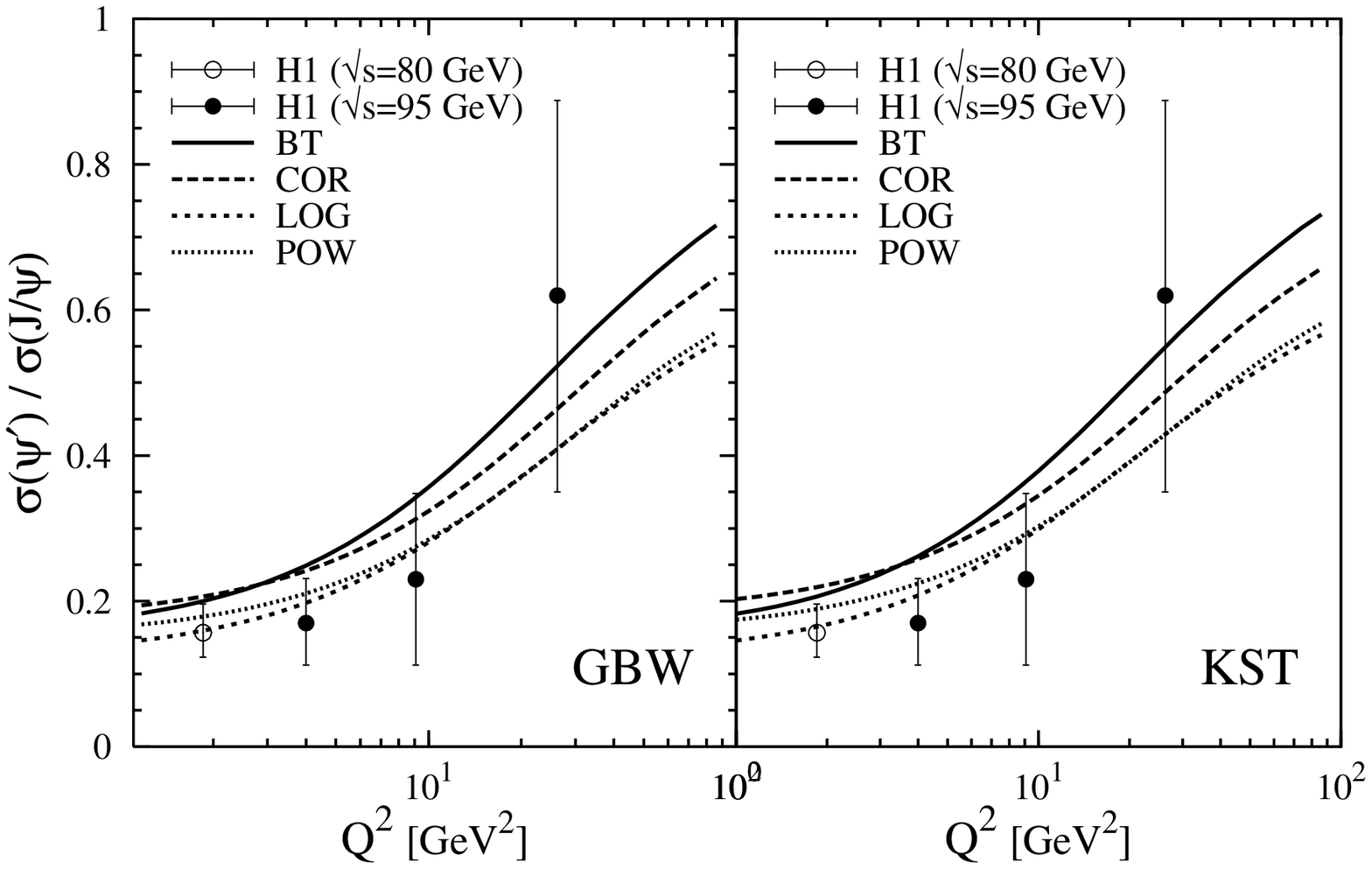}}}
\Caption{
  \label{Fig-RQ}
  The ratio of $\psi'$ to $J/\psi$ photoproduction cross sections as
  a function of the photon virtuality $Q^2$ at energy $\sqrt{s}=90\GeV$
  all four potentials with with GBW and KST parameterizations for the
  dipole cross section. Experimental data points from the H1 experiment
  \cite{H1-RQ}.
}
\EF

\section{Charmonium-nucleon total cross sections}\label{psi-n}

After the light-cone formalism has been checked with the data for virtual
photoproduction we are in position to provide reliable predictions for
charmonium-nucleon total cross sections. The corresponding expressions are
given by Eq.~\Ref{Mpsi}) (compare with \cite{ZKL}). For the GBW and KST
dipole cross sections, which have the form $\sigma_0\!\left(1-e^{-\rT^2%
/r_0^2}\right)$ (see Eqs.~\Ref{GBW} and \Ref{KST}), a summation over spin
indexes in \Ref{Mpsi} gives for the $S$-states,
\BE
  {\cal M}_{\psi\,p}(s) = \sigma_0 \cdot \left[
    1 - \pi r_0^2
    \int\limits_0^1     \!\!d\alpha
    \int\limits_0^\infty\!\!d\pT
    \int\limits_0^\infty\!\!d\qT
    \,U(\alpha,\pT) \,U(\alpha,\qT) \,V(\alpha,\pT,\qT)
    \right]\ ,
\EE
where
\BA
  U(\alpha,\pT)     &=& \pT\Phi_\psi(\alpha,\pT)
    \, e^{-r_0^2 \pT^2 / 4}
    \,\left[\Bigl(M_1^2(\pT)+\pT^2\Bigr)\Bigl(M_2^2(\pT)+\pT^2\Bigr)\right]^{-1/2}\ ,\\
  V(\alpha,\pT,\qT) &=&
            M_1(\pT) M_1(\qT) M_2(\pT) M_2(\qT)\, I_0(v) \\
  &+& \Bigl[M_1(\pT) M_1(\qT)+M_1(\pT) M_2(\qT)\Bigr]\,\pT \qT\, I_1(v)
   + \,\pT^2 \qT^2 \,I_2(v) \ ,\nonumber\\
  M_1(\pT) &=& m_c + m_T \sqrt{\frac{\alpha}{1-\alpha}} \ ,\\
  M_2(\pT) &=& m_c + m_T \sqrt{\frac{1-\alpha}{\alpha}}\ , \\
  v &=&{1\over2}\,r_0^2\,p_T\,q_T  \ .
\EA
Here $m_T^2=m_c^2+\pT^2$; $\Phi_\psi(\alpha,\pT)$ is defined in
(\ref{lc-wf-p}); $I_{0,1,2}(v)$ are Bessel functions of imaginary argument.

The calculated $\Jpsi$- and 
$\psi'$-nucleon total cross sections 
are plotted in Fig.~\ref{Fig-S} for for the GBW
and KST forms of the dipole cross sections and all four types of
the charmonium potentials. 

\BF
\centerline{\scalebox{0.94}{\includegraphics{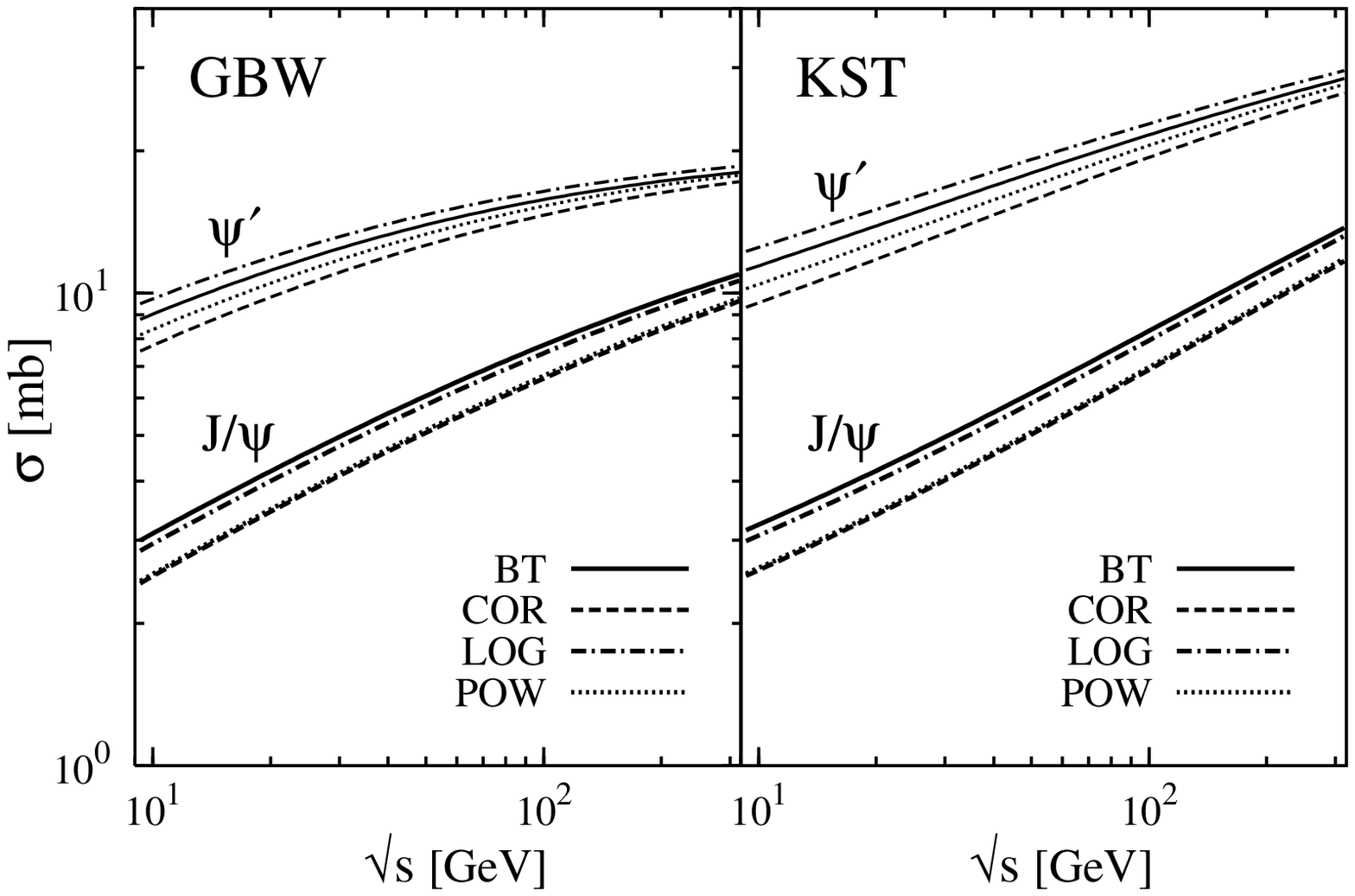}}}
\Caption{
  \label{Fig-S}
  Total $\Jpsi\,p$ (thick curves) and $\psi'\,p$ (thin curves)
  cross sections with the GBW and KST parameterizations for the
  dipole cross section.
}
\EF

The corresponding results for
$\chi$-states are depicted in Fig.~\ref{Fig-P}. 
\BF
\centerline{\scalebox{0.94}{\includegraphics{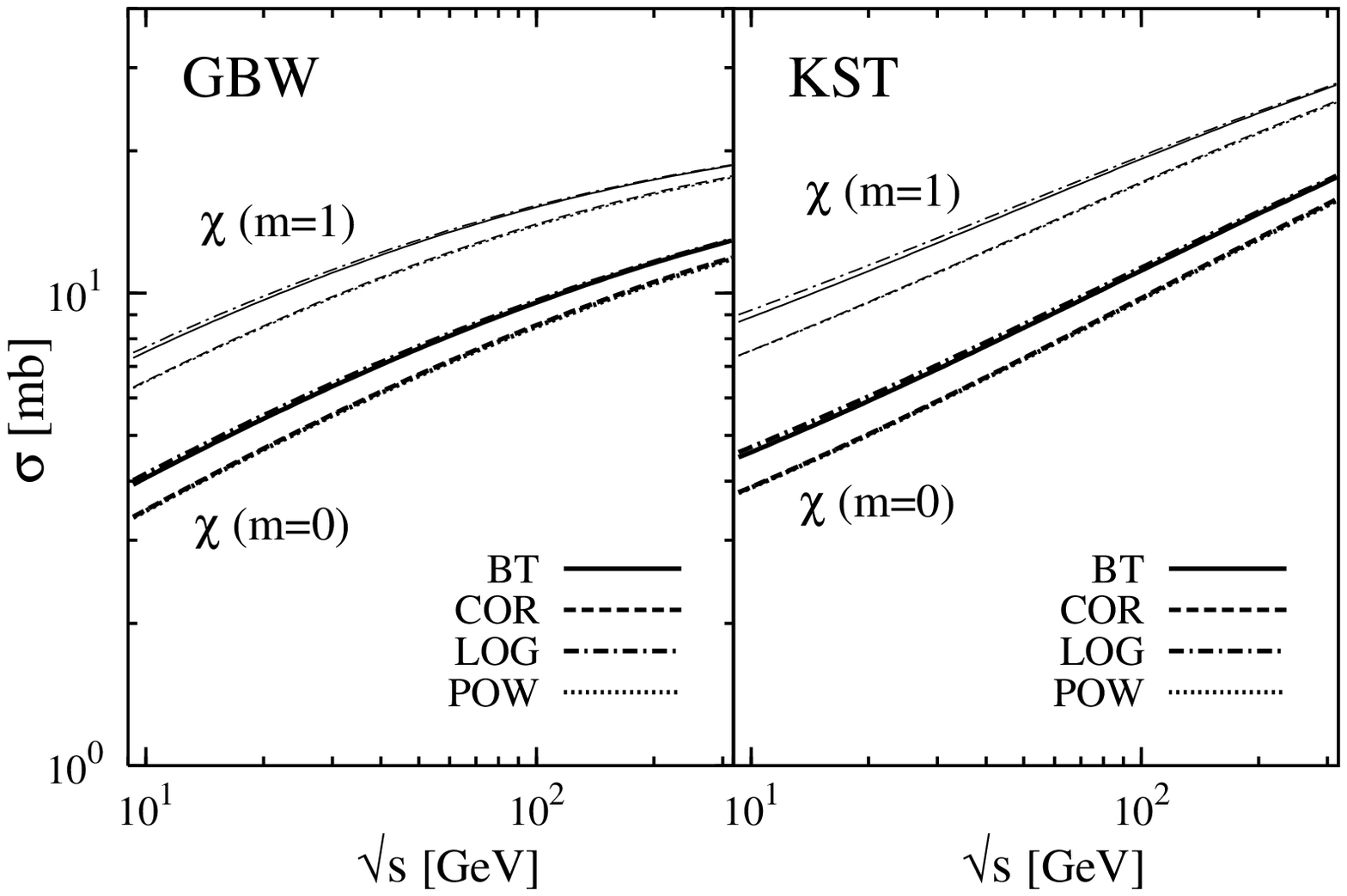}}}
\Caption{
  \label{Fig-P}
  Total $\chi\,p$ ($m=0$ --- thick curves, $m=1$ --- thin curves) cross
  sections with the KST and GBW parameterizations for the dipole cross
  section.
}
\EF
Here $m$ is the projection of the orbital momentum which can be zero or one, 
since this is a $P$-wave state. From these cross sections with definite $m$,
which we denote $\sigma^{\chi}_m$, one can construct the total cross sections
for the $\chi_c$ states with different spins and helicities $\lambda$,
\BA
\chi_{c0}(\lambda=0):\ \ \ \sigma &=& {1\over3}\,\Bigl(2\,\sigma^{\chi}_1 +
\sigma^{\chi}_0\Bigr)\ ;\nonumber\\
\chi_{c1}(\lambda=0):\ \ \ \sigma&=&\sigma^{\chi}_1\ ;\nonumber\\
\chi_{c1}(\lambda=\pm 1):\ \ \ \sigma &=& {1\over2}\,\Bigl(\sigma^{\chi}_1 +
\sigma^{\chi}_0\Bigr)\ ;\nonumber\\
\chi_{c2}(\lambda=0):\ \ \ \sigma &=& {1\over3}\,\,\Bigl(\sigma^{\chi}_1 +
2\,\sigma^{\chi}_0\Bigr)\ ;\nonumber\\
\chi_{c2}(\lambda=\pm 1):\ \ \ \sigma &=& {1\over2}\,\,\Bigl(\sigma^{\chi}_1 +
\sigma^{\chi}_0\Bigr)\ ;\nonumber\\
\chi_{c2}(\lambda=\pm 2):\ \ \ \sigma &=& \sigma^{\chi}_1\ .
\label{chi}
\EA
Using these relations one can easily derive the cross sections averaged
over helicities which are equal for all three states $\chi_{c\,0,1,2}$.

The strong dependence of the cross sections for $P$-wave charmonium states
on the projection $m=0,\ 1$ of the orbital momentum has been found
previously in \cite{fr}. However the predicted cross sections at
$\sqrt{s}=10\,GeV$ for
$\chi_c(m=0)$,  $\chi_c(m=1)$ and $\psi'$ are about twice as large as ours.
We believe that the disagreement originates from the too rough nonperturbative
dipole cross section\footnote{we are thankful to  Lars
Gerland who provided us with the expression for the dipole cross
section used in \cite{fr}.} used in \cite{fr} 
which was not well adjusted to data.
Even the pion-nucleon cross section calculated with Eq.~(1) in \cite{fr}
overestimates the experimental value by factor 1.5.

Although all four potentials are presented, comparison
with photoproduction data in Figs.~\ref{Fig-s} and \ref{Fig-QM} show
that two of them, BT and LOG potentials, are more trustable at least
for $\Jpsi$. These two potentials again give very close predictions for
$\Jpsi$-$p$ total cross sections but the deviation from the predictions
with the two other potentials, COR and POW, is much smaller than in
the case of photoproduction.

Note that the cross sections calculated with the GBW parameterization
demonstrate a tendency to level off at very high energy, especially for
$\psi'$, as compared to the KST predictions. The reason is obvious:
the GBW cross sections approach the universal limit
$\sigma_{max}=\sigma_0=23.03\,mb$. This cannot be true, and the KST
parameterization is more reliable than GBW at high energies where 
the gluon cloud surrounding the $\bar cc$ pair becomes nearly as big
as light hadrons.

According to Figs.~\ref{Fig-S} and \ref{Fig-P} for the KST parameterization
the total cross sections of charmonia are nearly straight lines as function of $\sqrt{s}$
in a double logarithmic representation, though with significantly different slopes for
the different states. Therefore a parameterization in the form
\BE
  \label{par-sD}
  \sigma^{\psi p}(s) =
  \sigma^{\psi}_0 \cdot \left(\frac{s}{s_0}\right)^{\Delta} \ ,
\EE
seems appropriate, at least within a restricted energy interval. We use the
data shown in Figs. \ref{Fig-S} and \ref{Fig-P} for the KST parameterization
of $\sigma_{q\bar q}$ and for the BT and LOG potentials and fit the them by
the form \Ref{par-sD} with $s_0=1000\GeV$. The two values from the BT and
LOG potentials have been averaged and their half difference gives the error
estimation. Table~\ref{Tab-sD} shows values for $\sigma^\psi_0$ and
$\Delta$ averaged over the energy interval $10\, GeV < \sqrt{s} < 300\, GeV$,
and the bound state sizes $\la \rT^2 \ra$. 
\BT
\Caption{
  \label{Tab-sD}
  Averaged sizes $\la\rT^2 \ra$ for charmonia bound states together 
  with $\sigma_0$ and $\Delta$ in the parameterization \Ref{par-sD} for
  the $\Jpsi$-, $\psi'$- and $\chi$-proton cross sections. Estimation of
  the errors is given in the text.
}
\vskip3mm
\begin{center}
\begin{tabular}{|c|c|c|c|}
\hline
\vphantom{\bigg\vert}
  & $\la \rT^2 \ra  \,[\fm^2\,]$ 
  & $\sigma^\psi_0\,[\mb  \,]$
  & $\Delta$ \\
\hline &&&\\[-4mm]
$\Jpsi$       & $0.117 \pm 0.003$ & $~5.59 \pm 0.13$ & $0.212 \pm 0.001$ \\
$\chi\,(m=0)$ & $0.181 \pm 0.004$ & $~7.17 \pm 0.07$ & $0.195 \pm 0.001$ \\
$\chi\,(m=1)$ & $0.362 \pm 0.007$ & $13.17 \pm 0.16$ & $0.164 \pm 0.002$ \\
$\psi'$       & $0.517 \pm 0.034$ & $16.63 \pm 0.59$ & $0.139 \pm 0.005$ \\[1mm]
\hline
\end{tabular}
\end{center}
\ET
As expected
$\sigma^\psi_0$ rises monotonically with the size of the charmonium state,
and the cross section for $\psi'\,N$ is about three times larger
that for $\Jpsi$. This deviates from the $r^2$ scaling, since the mean
value $\la r^2\ra$ is 4 times larger for $\psi'$ than for $\Jpsi$.
 The exponent $\Delta$ which governs the
 energy dependence decreases monotonically with the size of the charmonium
 state, demonstrating the usual correlation between the dipole size
 and the steepness of energy dependence. The values of $\Delta$ are larger
than in soft interactions of light hadrons ($\sim 0.08$), but smaller than values reached
in DIS at high $Q^2$.

Our results at $\sqrt{s}=10\,GeV$ (the mean energy of charmonia
produced in the NA38/NA50 experiments at SPS, CERN),
\BA
\sigma^{\Jpsi}_{tot}(\sqrt{s}=10\,GeV) &=& ~3.56 \pm 0.08 \mb\ , 
\label{1s}\\
\sigma^{\psi'}_{tot}(\sqrt{s}=10\,GeV) &=& 12.19 \pm 0.61 \mb\ ,
\label{2s}
\EA
well agree with the cross sections extracted in \cite{HK} 
from photoproduction data employing the 
two-channel approximation, $2.8 \pm 0.12\,mb <
\sigma^{J/\psi}_{tot}(\sqrt{s}=10\,GeV) <
4.1 \pm 0.15\ mb$ and $\sigma^{\psi'}_{tot}/\sigma^{J/\psi}_{tot} \approx
3.75$ (having poorly controlled accuracy), which shows that the two channel 
approach is a reasonable tool to analyze photoproduction data.

The cross section Eq.~(\ref{par-sD})) with the parameters in Table~\ref{Tab-sD}
agrees well with $\sigma^{\Jpsi}_{tot}(\sqrt{s}=20\,GeV)=4.4\pm0.6\,mb$
obtained in the model of the stochastic vacuum \cite{dosch}.

It worth noting that the results for charmonium-nucleon total cross
sections are amazingly similar to what one could get 
without any spin rotation,
\BE
  \label{Mpsi-nospin}
  \sigma_{tot}^{\psi\,N}(s) \approx
    \,\int\limits_0^1 \!\!d\alpha \int d^2\brT
    \,\left|\Phi_\psi(\alpha,\brT)\right|^2 \,\sigma_{q\bar q}(\rT,s)\ ,
\EE
where $\Phi_{\psi}(\alpha,\brT)$ is related by Fourier transformation to
Eq.~(\ref{lc-wf-p}), or even performing a simplest 
integration using the nonrelativistic wave
functions (\ref{Schroed}) in the rest frame of the charmonium:
\BE
  \label{Mpsi-NR}
  \sigma_{tot}^{\psi\,N}(s) \approx
  \int d^3\!r\,\left|\Psi(\vec r)\right|^2\,
  \sigma_{q\bar q}(\rT,s)\ .
\EE
The comparison presented in Fig.~\ref{Fig-Sc} for the BT potential
shows that (\ref{Mpsi-nospin}) - (\ref{Mpsi-NR}) are only about $10\%$
below the exact calculation for $J/\psi$, while there is practically no
difference between the exact and approximate calculations for $\psi'$.

\BF
\centerline{\scalebox{0.94}{\includegraphics{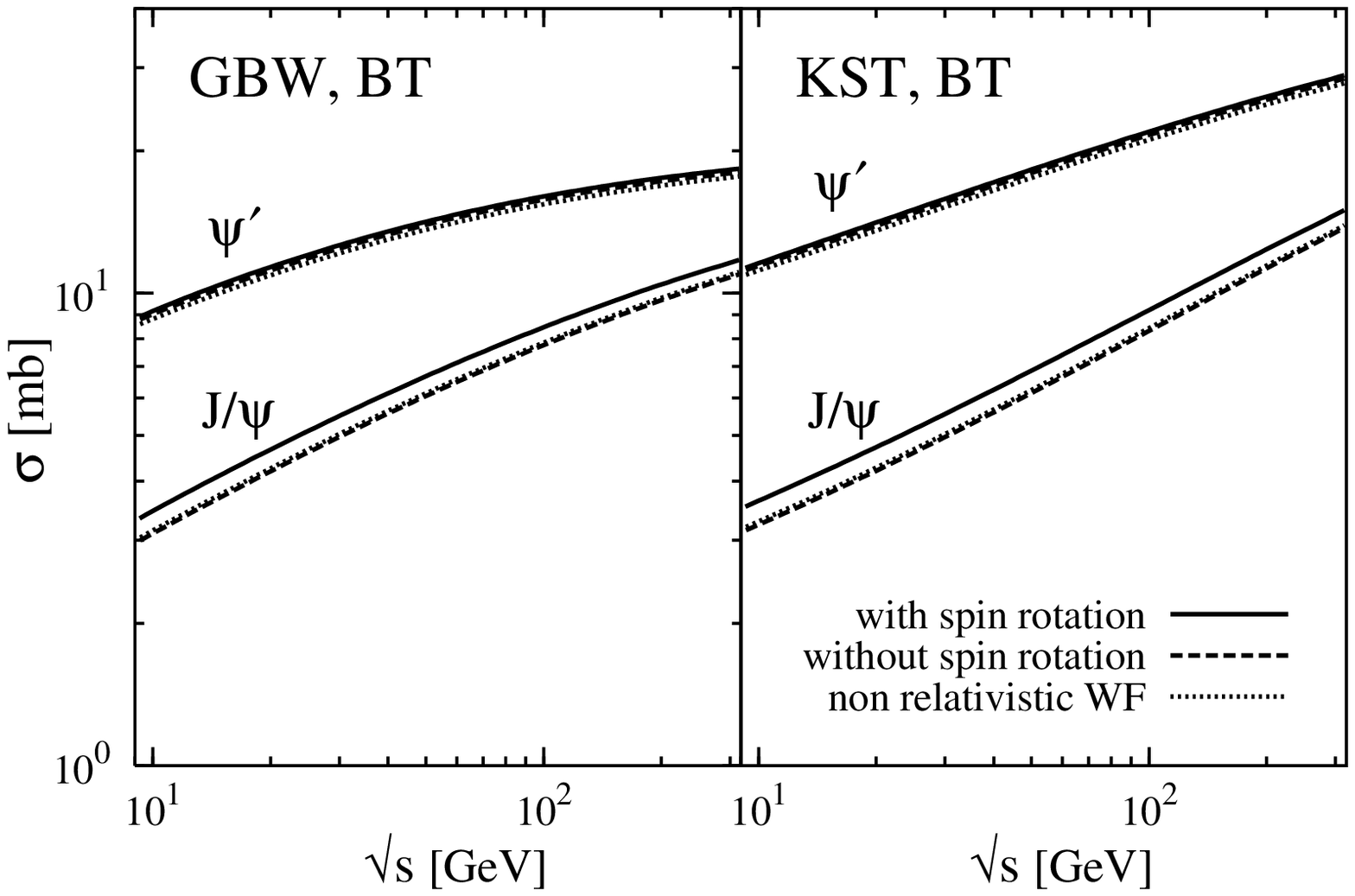}}}
\Caption{
  \label{Fig-Sc}
  Comparison of the results for $\sigma_{tot}^{\psi\,N}(s)$ obtained with
  the exact expression (\ref{Mpsi}) (solid curves) and with the approximations
  (\ref{Mpsi-nospin}) (dashed) and (\ref{Mpsi-NR}) (dotted).
}
\EF

\section{Nuclear suppression of charmonium production}\label{nuclei}

Production of charmonia off nuclei seem to be a natural source of information
about charmonium-nucleon cross section since nuclear absorption 
leads to suppression of the production rate measured experimentally.
However, one should be cautious applying our results to a calculation of nuclear
attenuation of charmonium. In exclusive photoproduction
of charmonia the uncertainty principle does not
allow to resolve between $J/\psi$ and $\psi'$ unless the formation
time, $t_f=2\,E_{\psi}/(M^2_{\psi'}-M^2_{J/\psi})$ \cite{KZ,hk-prl}, is shorter than
the mean inter-nucleon separation in nuclei. Only one experiment \cite{slac}
at $\sim 20\,GeV$ satisfies this condition. Analyzed with an optical model it
leads to $\sigma^{J/\psi\,N}_{in} = 3.5 \pm 0.8\,mb$ in a good agreement
with our calculations. The nuclear photoproduction 
data \cite{Sokolov} taken at $120\,GeV$ cannot be
treated in the same way since the formation time, $l_f\approx 10\,fm$
exceeds the nuclear size. In addition, the coherence length,
$l_c=2\,E_{\psi}/(M^2_{J/\psi}+Q^2)$ \cite{KZ,hkn}, is also long, about $5\,fm$
substantially increasing the attenuation path for the produced $\bar cc$ pair.

In the case of hadroproduction of charmonia off nuclei, the interplay
of the formation and coherence time effects are as important as in
photoproduction. On top of that, the situation is complicated
by decays of $\chi$s and $\psi'$ which substantially feed 
the yield of $J/\psi$.  These heavier states, even if their absorption
cross sections are known from our calculations, are also subject to
the effects of formation and coherence lengths.

In the analysis \cite{he} of data from the 
experiment E866 of $p A \to \Jpsi X$ collisions at 
$800\GeV$, proper attention has been given to coherence and formation
time effects with the result (extrapolated to $\sqrt s=10\GeV$)
\BA
  \sigma^{``\Jpsi'' p}_{eff} &=& ~5.0 \pm 0.4 \mb \ ,
  \label{he1}\\
  \sigma^{   \psi'  p}_{tot} &=& 10.5 \pm 3.6 \mb \ ,
\label{he2}
\EA

The effective $J/\psi$-nucleon cross section which is fed by decays of
heavier states can be estimated as follows,
\BE
\frac{1}{\sigma_{eff}}\,
\left(1 - e^{-\sigma_{eff}\,\la T\ra}\right) = 
\sum\limits_{i=1}^4\, 
\frac{w_i}{\sigma^{\psi_iN}_{tot}}\ \left(1 -
e^{-\sigma^{\psi_iN}_{tot}\,\la T\ra}\right)\ ,
\label{effective}
\EE
where $\la T\ra \approx 0.75\,\rho_A\,R_A$ is the mean thickness
of a nucleus with radius $R_A$ and the mean density $\rho_A\approx
0.16\,fm^{-3}$.

Eq.~(\ref{effective}) is relevant for $\Jpsi$ suppression in nuclear collisions (proton-%
nucleus and nucleus-nucleus). In this reactions the observed $\Jpsi$ arises
from directly produced $\Jpsi$'s with probability $w_1 < 1$ and from the
other states $\chi$, $\psi'$ via decay after the charmonia have left the
interaction zone, where $w_i$ is the probability that the state ``i'' contributes
to the finally observed $\Jpsi$. Values for $w_i$ and $\sigma^{\psi_i p}%
_{tot}$ are given in Table~\ref{Tab-Jeff} where $m=0,1$ is the projection of
the orbital momentum of the $\bar cc$ pair on the direction of gluon-gluon
collision in $\chi_{1,2}$ production ($\chi_0$ has a tiny branching to
$J/\psi$). 
\BT
\Caption{
  \label{Tab-Jeff}
  Values for the $\Jpsi$-, $\psi'$-, $\chi$- and effective ``$\Jpsi$''-proton
  cross sections at energy $\sqrt s=10\GeV$. Errors are given by averaging on
  BT and LOG potentials for the wave functions.
}
\vskip3mm
\begin{center}
\begin{tabular}{|c|c|c|}
\hline
\vphantom{\bigg\vert}
                 &    $w_i$   & $\sigma [\mb\,]$ \\ \hline &&\\[-4mm]
$\Jpsi$          & 0.52 - 0.6 & $~3.56 \pm 0.08$ \\
$\chi\,(m=0)$    & 0          & $~4.66 \pm 0.06$ \\
$\chi\,(m=1)$    & 0.32 - 0.4 & $~9.05 \pm 0.16$ \\
$\psi'$          & 0.08       & $12.19 \pm 0.61$ \\
\hline
\end{tabular}
\end{center}
\ET
It turns out that $\chi_1$ and $\chi_2$ with $m=0$ cannot be produced or are strongly
suppressed in gluon fusion
due to the selection rules which forbid projections $\pm 1$ for the total
angular momentum (e.g. see \cite{brodsky}), this is why we put $w_2=0$.

We calculate $\sigma_{eff}$ 
for tungsten used in the analysis \cite{he} and find for $\sqrt s=10\GeV$
\BE
  \sigma^{``\Jpsi''p}_{eff} = 5.8 \pm 0.2 \mb\ ,
\EE
where the main uncertainty arises from the $w_i$. This number is in a good
accord with Eq.~(\ref{he1}), while the calculated value for
$\sigma^{\psi'p}_{tot}$ Eq.~(\ref{2s}) agrees well with (\ref{he2}).

The coherence effects are quite important even at the energy of the
NA38/NA50 experiments ($E_{\psi}\approx 50\,GeV$) at CERN, this is why
the effective absorption cross section for $\psi'$ production suggested by
the data is about a half of the value we predict. At the energies of RHIC
and LHC both the coherence and formation times substantially exceed
the sizes of heavy nuclei, and shadowing becomes the dominant phenomenon.

\section{Conclusion and discussions}\label{summary}

In this paper we have proposed a simultaneous treatment of elastic
photoproduction $\sigma_{\gamma^*p \to \psi p}(s,Q^2)$ of charmonia
and total cross sections $\sigma^{\psi p}_{tot}(s)$. The ingredients
are (i) the factorized light-cone expressions (\ref{Mgam}) - (\ref{Mpsi})
for the cross sections;
(ii) the perturbative light-cone wave functions for the $c\bar c$ component of the
$\gamma^*$; (iii) light cone wave functions for the charmonia bound
states, and (iv) a phenomenological dipole cross section $\sigma
_{q\bar q}(\rT,s)$ for a $c\bar c$ interacting with a
proton.

The dipole cross section rises with energy; the smaller is
the transverse $\bar qq$ separation, the steeper is the growth. The source of
the energy dependence is the expanding cloud of gluons surrounding the $\bar qq$
pair. The gluon bremsstrahlung is more intensive for small dipoles. 
The gluon cloud can be treated as a joint
contribution of higher Fock states, $|\bar q\,q\,nG\ra$, however, it can
be also included into the energy dependence of $\sigma_{\bar qq}(r_T,s)$,
as we do, and this is the full description. Addition of any higher Fock state 
would be the double counting.

As function of energy the initial size of the
$\bar qq$ source is gradually ``forgotten'' after multi-step radiation,
the small cross sections grow steeper and eventually approach the larger ones
at very high energies.
All the cross sections are expected to reach a universal asymptotic 
behavior which saturates the Froissart bound.

The effective dipole cross section $\sigma_{q\bar q}(\rT,s)$ is parameterized
in a form which satisfies the expectations $\sigma_{q\bar q}\propto\rT^2$
for $\rT\to0$ (color transparency), but levels off 
for $\rT\to\infty$. Two parameterizations for $\sigma_{q\bar q}(\rT,s)$, whose
form and parameters have been fitted to describe $\sigma^{\pi p}_{tot}(s)$ and
the structure function $F_2(x,Q^2)$ are used in our calculations.

While the description of the photon wave function is quite certain,
the light-cone wave function of charmonia is rather ambiguous.
We have followed the usual recipe in going from a nonrelativistic wave
function calculated from a Schr\"odinger equation to a light cone form.
We have included the Melosh
spin rotation which is often neglected and found that it is instrumental 
to obtain agreement, since no parameter is adjustable.
In particular, it increases the $\psi'$ photoproduction
cross section by a factor  2 - 3 and rises the $\psi'$ to $J/\psi$ ratio to the
experimental value. 

At the same time, the charmonium-nucleon total cross sections ($J/\psi,\
\psi',\ \chi(m=0)$ and $\chi(m=1)$) turn out to be
rather insensitive to the way how the light-cone wave
function is formed, even applying no Lorentz transformation  one arrives at 
nearly the same results. This is why we believe 
that the predicted charmonium-nucleon cross section are very stable
against the ambiguities in the light-cone wave function of charmonia.
A significant energy dependence is predicted which varies from state to state
in accord with our expectations. 

We show our predictions for charmonium-nucleon cross sections
in a restricted energy range $10\,GeV < \sqrt{s} < 300\,GeV$,
but this interval can be largely extended in both directions.
Since the OZI rule suppresses the leading Reggeons, one can stay
with gluonic exchanges rather far down to low energies, unless
the charmed Reggeon exchanges become important \cite{barnes}.

\noindent {\bf Acknowledgment}: This work has been supported by a grant
from the Gesellschaft f\"ur Schwerionenforschung Darmstadt (GSI), grant
no. HD~H\"UFT and by the Federal Ministry BMBF grant no. 06 HD 954,
by the grant INTAS-97-OPEN-31696, and by the European Network: Hadronic
Physics with Electromagnetic Probes, Contract No. FMRX-CT96-0008


\end{document}